\documentclass[twocolumn,english,superscriptaddress]{revtex4}
\usepackage[T1]{fontenc}
\usepackage[latin1]{inputenc}
\usepackage{graphicx}

\makeatletter

\usepackage{babel}
\makeatother
\begin{document}

\title{Nonergodisity of a time series obeying L\'{e}vy statistics}

\author{Gennady Margolin}

\affiliation{Department of Chemistry and Biochemistry, Notre Dame University,
Notre Dame, IN 46556}

\author{Eli Barkai}

\affiliation{Department of Chemistry and Biochemistry, Notre Dame University,
Notre Dame, IN 46556}

\affiliation{Department of Physics, Bar Ilan University, Ramat Gan, Israel 52900}

\date{\today}

\begin{abstract}
Time-averaged autocorrelation functions of a dichotomous random process
switching between 1 and 0 and governed by wide power law sojourn time
distribution are studied. Such a process, called a L\'evy walk, describes
dynamical behaviors of many physical systems, fluorescence intermittency
of semiconductor nanocrystals under continuous laser illumination
being one example. When the mean sojourn time diverges the process
is non-ergodic. In that case, the time average autocorrelation function
is not equal to the ensemble averaged autocorrelation function, instead
it remains random even in the limit of long measurement time. Several
approximations for the distribution of this random autocorrelation
function are obtained for different parameter ranges, and favorably
compared to Monte Carlo simulations. Nonergodicity of the power spectrum
of the process is briefly discussed, and a nonstationary Wiener-Khintchine
theorem, relating the correlation functions and the power spectrum
is presented. The considered situation is in full contrast to the
usual assumptions of ergodicity and stationarity.
\end{abstract}
\maketitle

\section{Introduction}

Many time series exhibit a random behavior which can be represented
by a two-state process \cite{Allegrini-wrong}. In such processes
the state of the system will jump between state \emph{on} and state
\emph{off}. Examples include ion channel gating dynamics in biological
transport processes \cite{NadlerStein91,GoychukHanggi02} and gene
expression levels \cite{Dewey02,Roy} in cells, neuronal spike trains
\cite{Masuda}, motion of bacteria \cite{Korobkova04}, fluorescence
intermittency of single molecules \cite{Haase04} and nanocrystals
\cite{Nirmal,Kuno,Ken,Brokmann,Dahan}, and fluorescence fluctuations
of nanoparticles diffusing through a laser focus \cite{Zumofen04}.
Some aspects of spin dynamics can also be characterized using two
distinctive states \cite{GL,Bald}. These diverse systems may display
non-ergodicity and/or L\'{e}vy statistics \cite{BouchaudGeorges90,Schlesinger,MetzKlaf00,BJS04},
and often their behavior is found to deviate from simple scenarios
used in the past to interpret the behavior of ensembles. In particular,
in certain systems \cite{Nirmal,Kuno,Ken,Brokmann,Dahan,NadlerStein91,Korobkova04,Haase04,GL,Bald,Zumofen04}
power law sojourn times are found for one or both of the states. L\'{e}vy
statistics, which manifests itself in appearance of power laws, is
also found in flows on chaotic maps \cite{ZK}, which may be used
to model dynamics of various complex systems with non-linear interactions.
In this paper we address non-ergodicity of the L\'{e}vy walk processes
using a stochastic approach.

\begin{figure}
\includegraphics[%
  width=1.0\columnwidth,
  keepaspectratio]{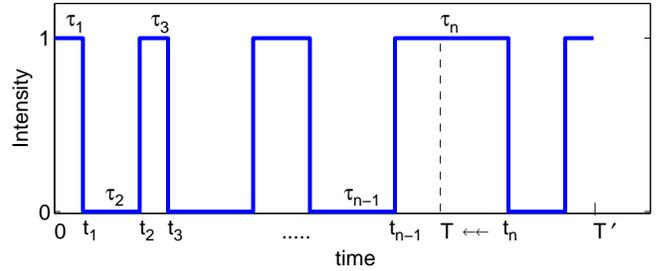}

\caption{\label{cap:schematic}Schematic representation of a dichotomous process.
$T=T'-t'$, where $T'$ is the duration of the experiment and $t'$
is the time difference used in correlation function (see Eq. (\ref{eq:Cst-def})).
Note that in Section \ref{sec:General-case} we redefine $t_{n}$
to be equal to \emph{T} and $\tau_{n}$ is redefined to be $T-t_{n-1}$,
to simplify notation.}
\end{figure}

We model the intermittent behavior by a random process which switches
between the two states after random sojourn times drawn from the probability
density functions (PDFs) $\psi_{\pm}(\tau)$, where the $\pm$ denote
the two states (see Fig. \ref{cap:schematic}).%
 It is assumed that these sojourn times are mutually independent random
variables. In the following we assume common PDF for both states $\psi(\tau)$,
unless stated otherwise, and assume that in state $+$, or \emph{on},
the system is described by the intensity $I=1$, while in state $-$,
or \emph{off}, it is described by zero intensity, $I=0$ (Fig. \ref{cap:schematic}).
We consider the case of power law decay for long times \begin{equation}
\psi(\tau)\sim\theta\tau^{-1-\theta},\,\,\,\,\,0<\theta<1,\label{eq:power-law-decay}\end{equation}
where we use natural units with dimensionless $\tau$. Such distributions
are observed in nanocrystal experiments \cite{Nirmal,Kuno,Ken,Brokmann,Dahan},
which under continuous laser illumination exhibit random two-state
blinking. As the mean sojourn time diverges, this situation reflects
aging and non-ergodicity. Aging means dependence of some observables
(e.g., ensemble average correlation functions) on absolute times from
the process onset at time zero, even in the limit of long times \cite{Marinari93,Chaos,Cheng,MB_JCP,Bouchaud92,LineShape04}.
Non-ergodicity means that ensemble averages are not equal to time
averages of single realizations, even in the limit of long times.

Generally speaking, our model represents the so-called L\'evy walk
process \cite{MetzKlaf00}, in which a particle travels on a line
with a constant velocity, changing directions at random times; the
sojourn times are distributed with a power-law decaying PDF $\psi(\tau)$.
Some of the systems mentioned above can in certain aspects be viewed
as physical realizations of the L\'evy walk.

In this manuscript we investigate the time average correlation function
of the L\'{e}vy walk process. When $\theta<1$ the process is nonergodic,
because the mean sojourn time diverges. It is a common practice to
replace the time average correlation function with the ensemble average
correlation function. Such a replacement is valid only for ergodic
processes. Previous attempt to model correlation function of the L\'{e}vy
walk process, ignored the problem of ergodicity \cite{Verberk}. Nonergodicity
was observed in experiments of Dahan's group \cite{Brokmann,Dahan},
who obtained nonergodic correlation functions in experiments on nanocrystals.
However, as far as we know there is no attempt to quantify the nonergodic
properties of correlation functions of blinking nano-crystals and
other L\'{e}vy walk processes. Such a quantification is important
in understanding the unusual behavior of physical systems and mathematical
models described in terms of L\'evy walks. Here we present a detailed
analysis of our findings, part of which was reported in \cite{MB_PRL}.

\section{Time average correlation functions}

We consider an \emph{on}-\emph{off} signal in the interval $(0,T')$
with intensity $I(t)$ jumping between two states $I(t)=1$ and $I(t)=0$.
At start of the measurement $t=0$ the process begins in state \emph{on}
$I(0)=1$. The process is characterized based on the sequence $\{\tau_{1}^{on},\tau_{2}^{off},\tau_{3}^{on},\tau_{4}^{off},\cdots\}$
of \emph{on} and \emph{off} sojourn times or equivalently according
to the dots on the time axis $t_{1},t_{2},\cdots$, on which transitions
from \emph{on} to \emph{off} or vice versa occur (cf. Fig. \ref{cap:schematic}).
Define the following time-averaged (TA) correlation function for a
single realization/trajectory:

\begin{equation}
C_{TA}(t',T')=\frac{\int_{0}^{T'-t'}I(t)I(t+t')dt}{T'-t'}=\frac{\int_{0}^{T}I(t)I(t+t')dt}{T},\label{eq:Cst-def}\end{equation}
and we denoted \[
T=T'-t'>0.\]
We are interested in the asymptotic behavior of the correlation function
for large \emph{T} and $t'$, and define a ratio\begin{equation}
r=\frac{t'}{T'},\label{eq:r-def}\end{equation}
which will be a useful parameter. In the non-ergodic situations we
consider, the distribution of the correlation function will asymptotically
depend on $t'$ and $T'$ only through their ratio \emph{r}.

The mathematical goal of this paper is to investigate the PDF of $C_{TA}(t',T')$.
We first consider the PDF of $C_{TA}(t',T')$ in the ergodic case,
and then address the non-ergodicity for $\theta<1$. This PDF is denoted
by $P_{C_{TA}(t',T')}(z)$, where $0\leq z\leq1$ are possible values
of $C_{TA}(t',T')$, due to Eq. (\ref{eq:Cst-def}).

\subsection{Ergodic case\label{sub:Ergodic-case}}

\begin{figure}
\includegraphics[%
  width=1.0\columnwidth]{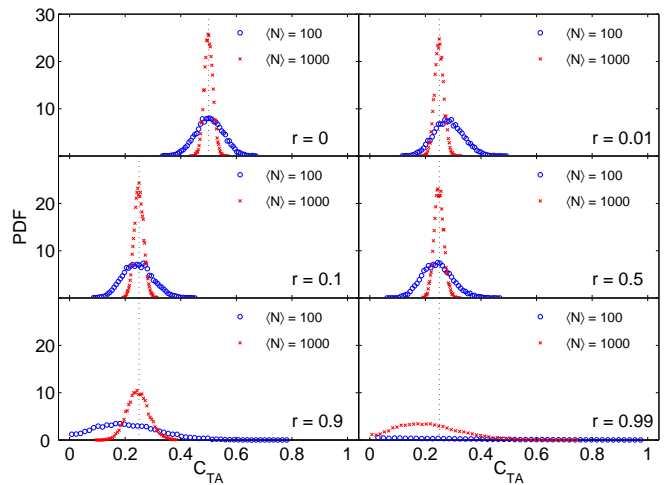}

\caption{\label{cap:exponential}Distribution of time-averaged correlation
function for $\psi(\tau)=e^{-\tau}$ is seen to approach the Dirac
delta function as the average number of transitions per realization
$\left\langle N\right\rangle $ grows. Location of the delta function
shifts from $1/2$ for $r=0$ to $1/4$ for any $r\neq0$ for large
enough $T'$(and hence also $t'$), as indicated by the dotted line.
Here $\left\langle N\right\rangle =T'$.}
\end{figure}

Let us first consider the ergodic case with exponential PDF of sojourn
times $\psi(\tau)=e^{-\tau}$, when the mean sojourn time defined
by \[
\left\langle \tau\right\rangle =\int_{0}^{\infty}\tau\psi(\tau)d\tau=1\]
 is finite. If the process is ergodic, the PDF of $C_{TA}(t',T')$
will approach in the limit of long times $T'\rightarrow\infty$, the
Dirac delta function\begin{equation}
P_{C_{TA}(t',T')}(z)\sim\delta\left(z-\left\langle C_{TA}(t',T')\right\rangle \right),\label{eq:ergodic-PDF}\end{equation}
where $\left\langle \right\rangle $ represent ensemble average. This
is what we mean by ergodicity of the two-time correlation function.
We illustrate this behavior in Figure \ref{cap:exponential}, using
numerical simulations. Increasing the experimental time $T'$ (and
hence also $t'$, to keep \emph{r} constant) leads to narrowing of
the distribution of the correlation function, yielding asymptotically
Eq. (\ref{eq:ergodic-PDF}). It is also clear that, for any nonzero
\emph{r} the ensemble average $\left\langle C_{TA}(t',T')\right\rangle $
will tend to $(1/2)^{2}=1/4$ as we increase $T'$. Stretching of
the distributions observed in Fig. \ref{cap:exponential} for large
\emph{r} is due to the finiteness of $T'$: here $T=T'-t'$ becomes
of the order of unity, which is the mean time of $e^{-\tau}$. Therefore,
this behavior is completely pre-asymptotic.

The picture is completely different when we consider Eq. (\ref{eq:power-law-decay})
with $\theta<1$, as is shown below. There is no narrowing of the
distribution, and it actually tends to a universal shape, which is
a function of \emph{r} and $\theta$ alone. The analogue of this distribution
in the ergodic case is the Dirac delta, Eq. (\ref{eq:ergodic-PDF}).
In the ergodic case, one is usually interested in the non-universal
behavior for relatively short $t'$ of the order of mean sojourn time,
while for long $t'$ the behavior is trivial. On the contrary, in
the non-ergodic regime we consider, the behavior of interest in this
paper is the universal nontrivial asymptotic behavior. From now on,
$\theta<1$ \cite{1<theta<2}.

\subsection{Non-ergodic case}

\begin{figure}
\includegraphics[%
  width=1.0\columnwidth,
  keepaspectratio]{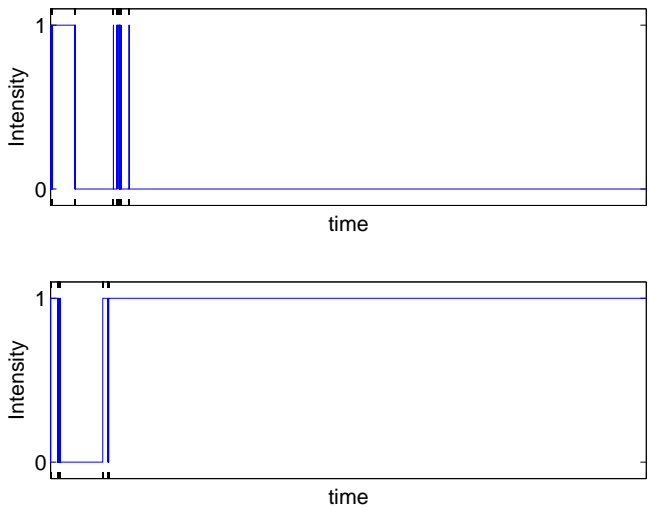}

\caption{\label{cap:Two-traj}Two randomly selected trajectories for $\theta=0.3$.
There are approximately 1000 transitions in each trajectory. The behavior
is dominated by a few large intervals and hence is strongly nonergodic.}
\end{figure}

\begin{figure}
\includegraphics[%
  width=85mm,
  height=35mm]{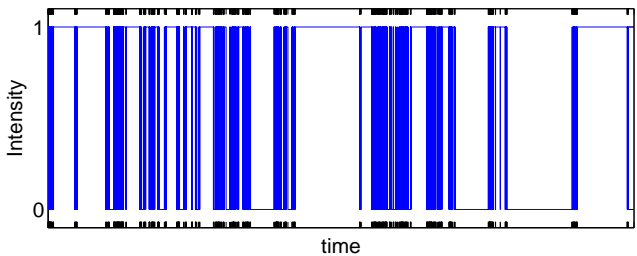}

\caption{\label{cap:One-traj.8}One randomly selected trajectory for $\theta=0.8$
with 1000 transitions. In comparison to $\theta=0.3$ (Fig. \ref{cap:Two-traj}),
the longest sojourn times here are shorter and the behavior is less
nonergodic.}
\end{figure}

\begin{figure}
\includegraphics[%
  width=1.0\columnwidth,
  keepaspectratio]{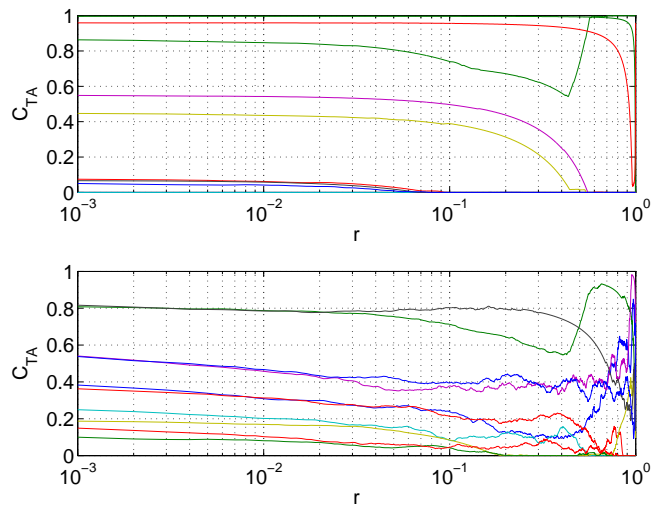}

\caption{\label{cap:Cta-realiz}Ten typical realizations of $C_{TA}$ dependence
on \emph{$r=t'/T'$} for $\theta=0.3$ (top) and $\theta=0.8$ (bottom).
$T'$ is kept constant, $t'$ changes. For an ergodic process all
correlation functions would follow the same master curve, the ensemble
average correlation function.}
\end{figure}

We begin the discussion of a non-ergodic situation by illustrating
two randomly selected trajectories for $\theta=0.3$ in Fig. \ref{cap:Two-traj}.%
 Clearly, these two trajectories are different, and hence time averaged
correlation functions of these two trajectories will be different,
yielding ergodicity breaking. It is important to emphasize that increasing
the measurement time $T'$, would not yield an ergodic behavior, since
the process has no characteristic average time scale. In Fig. \ref{cap:One-traj.8}%
 we show one trajectory with $\theta=0.8$ to compare to Fig. \ref{cap:Two-traj}.
One can say that for $\theta=0.8$ the nonergodicity is weaker. Unlike
Fig. \ref{cap:Two-traj}, in Fig. \ref{cap:One-traj.8} we do not
see one long \emph{on} or \emph{off} period dominating the time series.
In Figure \ref{cap:Cta-realiz} we plot ten typical realizations of
a correlation function, for a power-law decaying $\psi(\tau)$ following
Eq. (\ref{eq:power-law-decay}) with $\theta=0.3$ and $\theta=0.8$.
The most striking feature of this figure is that the correlation functions
are random. For very small \emph{r} there is more or less smooth evolution
of the correlation functions. As \emph{r} grows their behavior becomes
more chaotic.%
 We stress that this randomness is a true behavior and is \emph{not}
a problem in our simulations.

\begin{figure}
\includegraphics[%
  width=1.0\columnwidth,
  height=1.0\columnwidth,
  keepaspectratio]{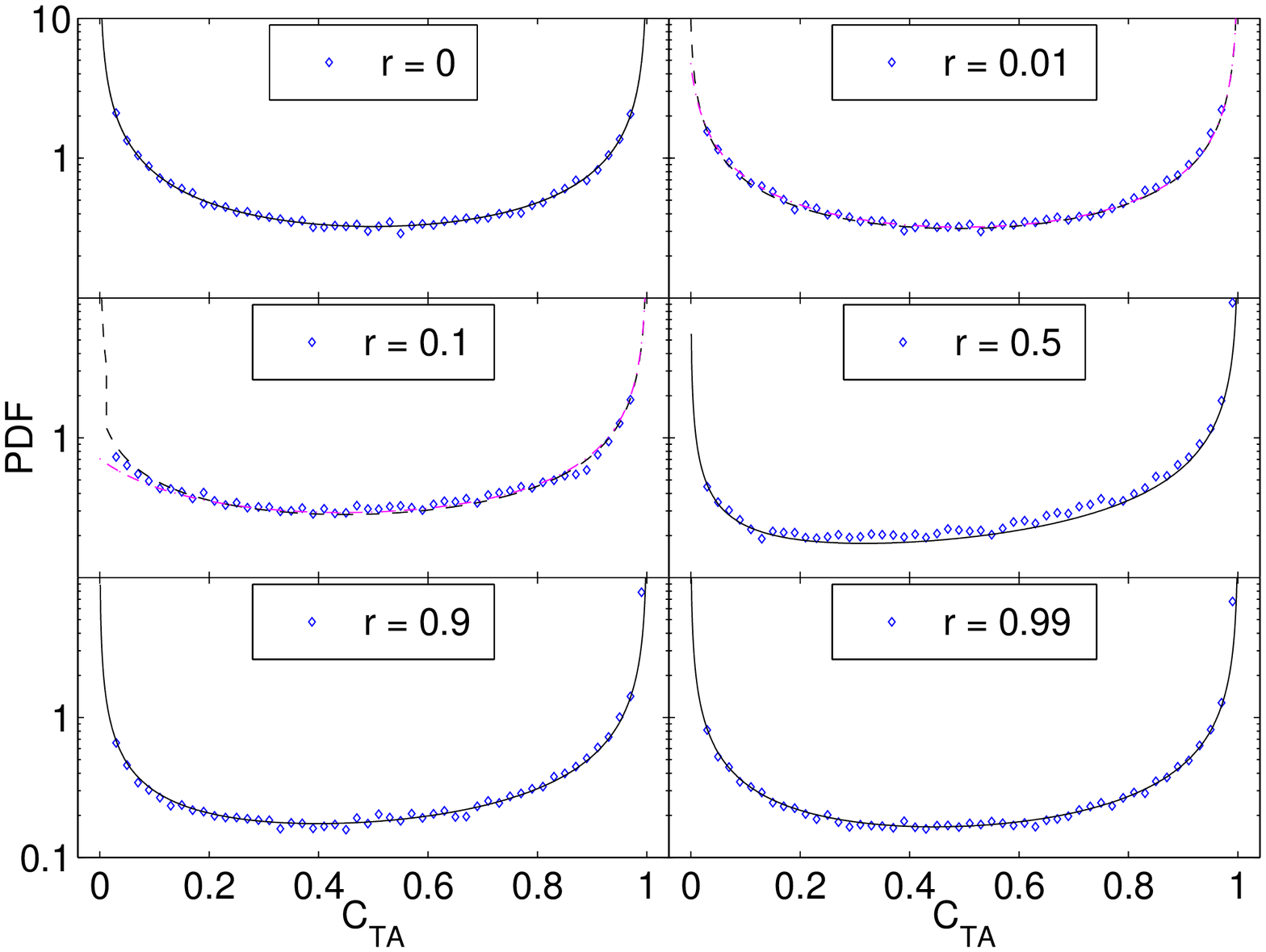}

\caption{\label{cap:theta0.3}PDF of $C_{TA}(t',T')$ for different $r=t'/T'$
and $\theta=0.3$. $\left\langle N\right\rangle \approx10^{3}$, $T'\approx1.66\times10^{10}$.
Abscissas are possible values of $C_{TA}(t',T')$. Diamonds are numerical
simulations. Curves are analytical results without fitting: for $r=0$
Eq. (\ref{eq:lamperti}) is used (full line), for $r=0.01$ and 0.1
Eq. (\ref{eq:PDF-Cst-t'<<T}) is used (dashed) and for $r=0.5$, 0.9
and 0.99 Eq. (\ref{eq17}) is used (full). See Section \ref{sec:Numerical-simulations}
for details.}
\end{figure}

\begin{figure}
\includegraphics[%
  width=1.0\columnwidth,
  height=1.0\columnwidth,
  keepaspectratio]{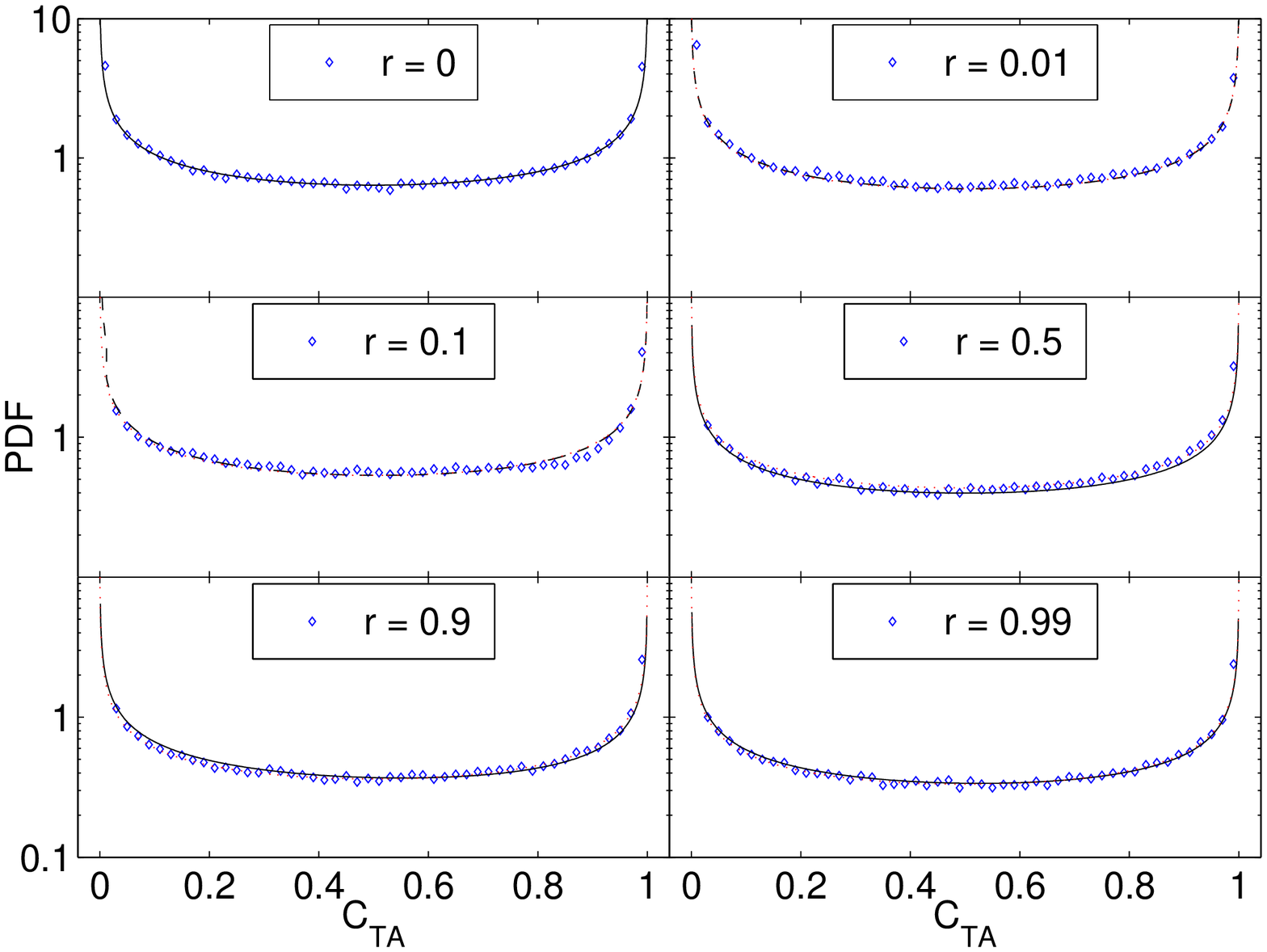}

\caption{\label{cap:theta0.5}PDF of $C_{TA}(t',T')$ for different $r=t'/T'$
and $\theta=0.5$. $\left\langle N\right\rangle \approx10^{3}$, $T'\approx2.47\times10^{6}$.
Diamonds are numerical simulations. Curves are analytical results
without fitting: for $r=0$ Eq. (\ref{eq:lamperti}) is used (full
line), for $r=0.01$ and 0.1 Eq. (\ref{eq:PDF-Cst-t'<<T}) is used
(dashed) and for $r=0.5$, 0.9 and 0.99 Eq. (\ref{eq17}) is used
(full). See Section \ref{sec:Numerical-simulations} for details.}
\end{figure}

\begin{figure}
\includegraphics[%
  width=1.0\columnwidth,
  height=1.0\columnwidth,
  keepaspectratio]{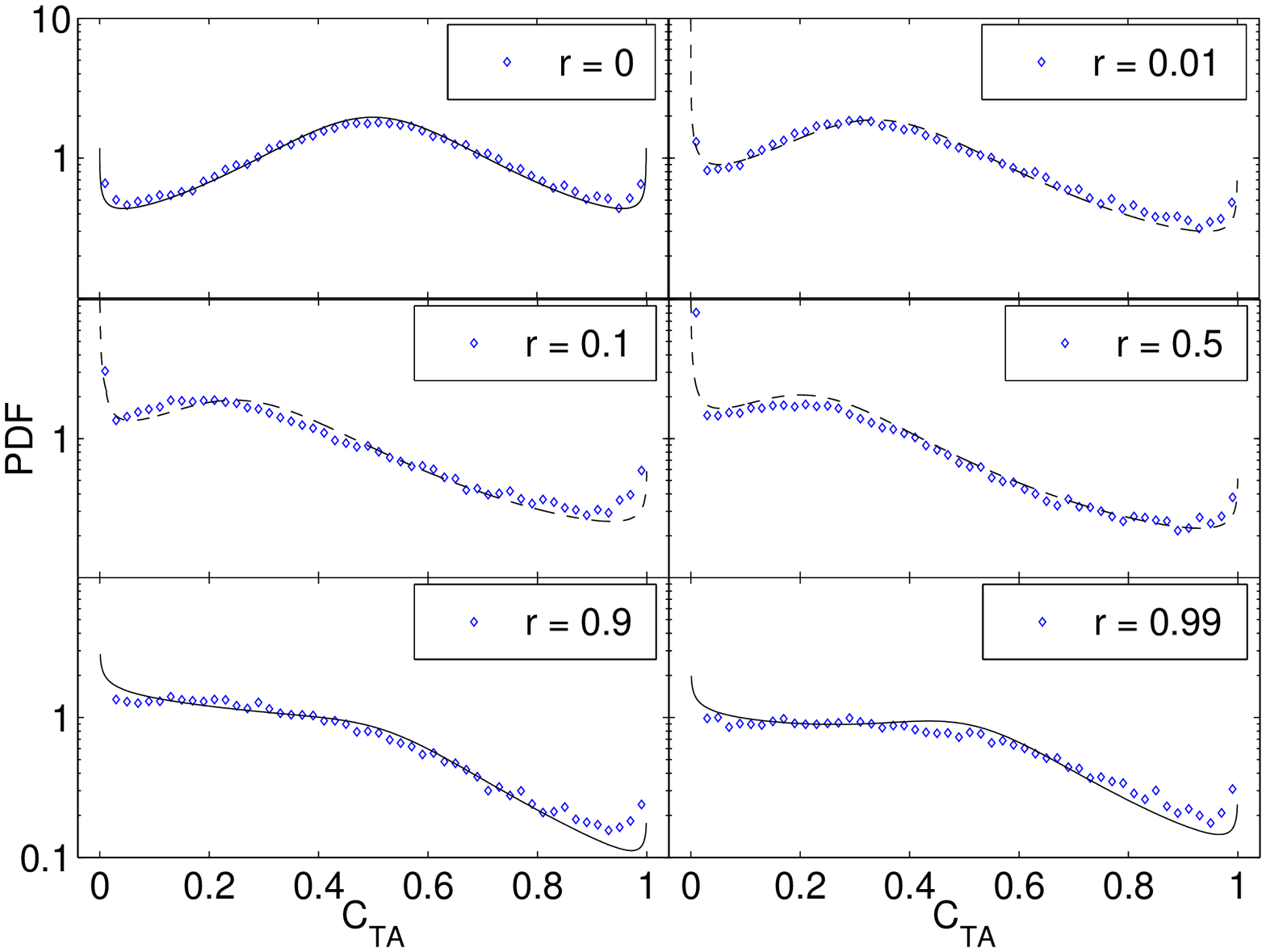}

\caption{\label{cap:theta0.8}PDF of $C_{TA}(t',T')$ for different $r=t'/T'$
and $\theta=0.8$. $\left\langle N\right\rangle \approx10^{4}$, $T'\approx6.15\times10^{5}$.
Diamonds are numerical simulations. Curves are analytical results
without fitting: for $r=0$ Eq. (\ref{eq:lamperti}) is used (full
line), for $r=0.01$, 0.1 and 0.5 Eq. (\ref{eq:PDF-Cst-t'<<T}) is
used (dashed) and for $r=0.9$ and 0.99 Eq. (\ref{eq17}) is used
(full). See Section \ref{sec:Numerical-simulations} for details.}
\end{figure}

For many realizations, our numerical simulations are used to obtain
$P_{C_{TA}(t',T')}(z)$ depicted in Figures \ref{cap:theta0.3}%
, \ref{cap:theta0.5} and \ref{cap:theta0.8} for $\theta=0.3$, $\theta=0.5$
and $\theta=0.8$, respectively ($\left\langle N\right\rangle $ is
the average number of transitions per realization; details of these
simulations are deferred until Section \ref{sec:Numerical-simulations}
and theoretical analysis is developed in Section \ref{sec:General-case}
below). The diamonds are numerical results. In all the figures we
vary $r\equiv t'/T'$. First consider the case $r=0$. For $\theta=0.3$
and $\theta=0.5$ we see from Figs. \ref{cap:theta0.3} and \ref{cap:theta0.5}
that the PDF $P_{C_{TA}(t',T')}(z)$ has a $U$ shape. This is a strong
non-ergodic behavior, since the PDF does not peak on the ensemble
averaged value of the correlation function which is $1/2$. On the
other hand, when $\theta=0.8$ the PDF $P_{C_{TA}(t',T')}(z)$ has
a $W$ shape (cf. Fig. \ref{cap:theta0.8}), a weak non ergodic behavior.
To understand the origin of this type of transition note that as $\theta\rightarrow0$
we expect the process to be in an \emph{on} state or an \emph{off}
state for the whole duration of the measurement. This is so because
the probability that the sojourn time is longer then $T'$ will be
$\sim(T')^{-\theta}\rightarrow1$ (cf. Fig. \ref{cap:Two-traj}).
Hence in that case the PDF of the correlation function will peak on
$C_{TA}(t',T')=1$ and $C_{TA}(t',T')=0$ (i.e $U$ shape behavior).
On the other hand when $\theta\rightarrow1$ we expect a more ergodic
behavior, since for $\theta>1$ the mean \emph{on} and \emph{off}
periods are finite, this manifests itself in a peak of the distribution
function of $C_{TA}(t',T')$ on the ensemble average value of $1/2$
and a $W$ shape PDF emerges (Fig \ref{cap:theta0.8}, $r=0$). Note
that for $\theta<1$ there is still statistical weight for trajectories
which are \emph{on} or \emph{off} for periods of the order of the
measurement time $T'$, and the distribution of $C_{TA}(0,T')$ attains
its maximum on $C_{TA}(0,T')=1$ and $C_{TA}(0,T')=0$. 

For $r>0$ we observe in Figs. \ref{cap:theta0.3} and \ref{cap:theta0.8}
non-symmetrical and non-trivial shapes of the PDF of the correlation
function. These PDFs agree very well with the analytical results,
which we derive later. Not shown in Figs. \ref{cap:theta0.3}, \ref{cap:theta0.5}
and \ref{cap:theta0.8} is a delta function contribution on $C_{TA}(t',T')=0$.
In other words, for $t'\neq0$, some of the random correlation functions
are equal zero. The number of such correlation functions is increasing
when \emph{r} is increased. When $r\rightarrow1$, half of the correlation
functions are equal to zero (see Section \ref{sec:General-case}).
Qualitatively, considering large \emph{r}, the correlation is between
the signal close to its starting point and the signal close to its
end point. Roughly speaking, close to the end of the signal, typically
long sojourn intervals with no transitions occur (cf. Fig. \ref{cap:Two-traj};
i.e. persistence, as explained later in the paper in more detail -
cf. Eq. (\ref{eq:p0-def})). For those types of trajectories being
in state \emph{off} at the end, the correlation function should be
zero. We stress that the distributions observed on Figs. \ref{cap:theta0.3},
\ref{cap:theta0.5} and \ref{cap:theta0.8} are not a scaling artifact:
analogous calculations in the case of $\theta>1$ lead in the limit
$T'\rightarrow\infty$ to Dirac $\delta$-functions instead, as was
shown above (Section \ref{sub:Ergodic-case}; cf. Fig. \ref{cap:exponential}). 


We now turn to an analytical treatment of the described non-ergodicity.

\section{$t'=0$: Lamperti distribution}

In the case $t'=0$ there exists known asymptotically exact expression
for $P_{C_{TA}(0,T)}(z)$. Let us define \begin{equation}
\mathcal{I}_{[a,b]}=\frac{\int_{a}^{b}I(t)\textrm{d}t}{b-a},\label{eq:Iab}\end{equation}
the time average intensity between time $a$ and time $b>a$. For
$t'=0$ from Eq. (\ref{eq:Cst-def}) it immediately follows that the
time averaged correlation function is identical to the time average
intensity \begin{equation}
C_{TA}(0,T)=\mathcal{I}_{[0,T]}=\frac{T_{[0,T]}^{+}}{T}\label{eq:Cta-TplusoverT}\end{equation}
where $T_{[a,b]}^{+}$ is the total time \emph{on} of a particular
realization in the time interval $[a,b]$. The time average intensity
$\mathcal{I}_{[0,T]}$ has a known asymptotic distribution as $T\rightarrow\infty$,
found originally by Lamperti \cite{Lamp,GL} and denoted in this paper
as $\ell_{\theta}$:\[
P_{C_{TA}(0,T)}(z)=l_{\theta}\left(z\right),\]
and \begin{equation}
l_{\theta}\left(z\right)=\frac{\sin\pi\theta}{\pi}\frac{z^{\theta-1}\left(1-z\right)^{\theta-1}}{z^{2\theta}+(1-z)^{2\theta}+2z^{\theta}\left(1-z\right)^{\theta}\cos\pi\theta},\label{eq:lamperti}\end{equation}
for $0\leq z\leq1$. For negative \emph{z} and for $z>1$ it is zero.
Note that $\ell_{\theta}(z)=\ell_{\theta}(1-z)$ and $\ell_{\theta}(z)$
diverges at $z=0,1$. This function is normalized to 1 for any $0<\theta\leq1$.
The Lamperti PDF is shown in Figs. \ref{cap:theta0.3}, \ref{cap:theta0.5}
and \ref{cap:theta0.8} for the case $r=0$, together with the numerical
results. The transition between the $U$ shape behavior and the $W$
shape behavior happens at $\theta_{c}=0.5946...$. Lamperti distribution
is related to the well known arcsine law \cite{Feller} (case $\theta=1/2$).
Other works regarding relative time spent by a system in one of two
states are \cite{Bald,Dhar,BelBarkai}.

\section{Ensemble average $\left\langle C_{TA}(t',T')\right\rangle $\label{sec:Ensemble-average}}

Another useful asymptotically exact result that can be derived is
the mean of $P_{C_{TA}(t',T')}(z)$, i.e., the ensemble average of
$C_{TA}(t',T')$. Generalizing to slightly different \emph{on} and
\emph{off} time PDFs, with equal exponents but different coefficients,\begin{equation}
\psi_{\pm}(t)\sim A_{\pm}t^{-1-\theta},\label{eq:psi+-}\end{equation}
it has been shown \cite{MB_JCP} that the mean intensity-intensity
correlation function is asymptotically, for $t'\geq0$\begin{equation}
\left\langle I(t)I(t+t')\right\rangle \sim P_{+}-P_{+}P_{-}\frac{\sin\pi\theta}{\pi}B\left(\frac{1}{1+t/t'};1-\theta,\theta\right),\label{eq:I-I-average}\end{equation}
where the incomplete beta function is defined as

\begin{equation}
B(z;\alpha,\beta)=\int_{0}^{z}x^{\alpha-1}(1-x)^{\beta-1}dx\label{eq:Beta}\end{equation}
and\[
P_{\pm}=\frac{A_{\pm}}{A_{+}+A_{-}}.\]
In the particular case of equal $\psi_{\pm}(t)$ we have $P_{\pm}=1/2$.
Eq. (\ref{eq:I-I-average}) exhibits aging since the correlation function
depends on \emph{t} even when it is long. Aging of the ensemble average
correlation function is related to nonergodicity of single realization
trajectory.

Integrating we thus obtain from Eq. (\ref{eq:Cst-def})\begin{equation}
\begin{array}{c}
\left\langle C_{TA}(t',T')\right\rangle ={\displaystyle \frac{\int_{0}^{T}\left\langle I(t)I(t+t')\right\rangle dt}{T}}\sim P_{+}^{2}+P_{+}P_{-}\times\\
\\{\displaystyle \frac{\sin\pi\theta}{\pi}}\left[{\displaystyle \frac{B(1-r;\theta,1-\theta)}{1-r}-\frac{1}{\theta}\left(\frac{r}{1-r}\right)^{1-\theta}}\right].\end{array}\label{eq:meanCst-theor}\end{equation}
We see that the mean of the single trajectory correlation function
asymptotically depends only on the ratio \emph{r} of its arguments.
We will show that the same is true also for the whole PDF of this
random function, and not only for its mean. For \emph{r} close to
zero and to one,\begin{equation}
\begin{array}{c}
\left\langle C_{TA}(t',T')\right\rangle \sim\qquad\qquad\qquad\qquad\qquad\\
\\\left[\begin{array}{cc}
P_{+}\left(1-(1-P_{+}){\displaystyle \frac{r^{1-\theta}\sin\pi\theta}{\pi\theta(1-\theta)}}\right), & r\ll1\\
\\P_{+}^{2}+P_{+}P_{-}{\displaystyle \frac{(1-r)^{\theta}\sin\pi\theta}{\pi\theta(1+\theta)}}, & 1-r\ll1.\end{array}\right.\end{array}\label{eq:Cta-ens-av}\end{equation}

It is worth mentioning that for an \emph{ergodic} time series the
variance \[
\sigma_{\mathcal{I}}^{2}(T)=\left\langle \left(\mathcal{I}_{[0,T]}-\left\langle \mathcal{I}_{[0,T]}\right\rangle \right)^{2}\right\rangle =\left\langle \mathcal{I}_{[0,T]}^{2}\right\rangle -\left\langle \mathcal{I}_{[0,T]}\right\rangle ^{2}\]
 should go to zero as $T\rightarrow\infty$. In the case $\theta<1$,
in this limit $\left\langle \mathcal{I}_{[0,T]}\right\rangle \rightarrow P_{+}$
\cite{MB_JCP} and using Eq. (\ref{eq:I-I-average}),\begin{equation}
\sigma_{\mathcal{I}}^{2}(T)\rightarrow\frac{\sin\pi\theta}{\pi}P_{+}P_{-}\int_{0}^{1}B(x;\theta,1-\theta)dx=P_{+}P_{-}(1-\theta),\label{eq:var-theta}\end{equation}
which is non-zero, and so we can prove the non-ergodicity of the considered
process, even without knowing $P_{C_{TA}(t',T')}(z)$. The last equality
can be easily obtained using Eq. (\ref{eq:Beta}). 

We conclude this section by introducing the probability $p_{0}(a,b)$
of making no transition, either up to down or vice versa, between
two arbitrary times $a$ and $b\geq a$, known as the persistence
probability. For large $a$ (cf. Eq. (\ref{eq:p0-int-fE}))\begin{equation}
p_{0}(a,b)\sim\frac{\sin\pi\theta}{\pi}B\left(a/b;\theta,1-\theta\right).\label{eq:p0-def}\end{equation}
Without going into details, we note that this probability plays important
role in L\'evy walks, and in particular in formulas given above \cite{GL,MB_JCP}.
Its crucial feature is that it depends on the ratio of times and not
on their difference, as is the case for ergodic processes. See also
\cite{Dhar,Majumdar}.

\textbf{Remark:} Eq. (\ref{eq:var-theta}) also follows from the fact
that $\sigma_{\mathcal{I}}^{2}(T)$ should approach the variance of
the Lamperti distribution (for $P_{+}=P_{-}$), whose moments can
be calculated \cite[appendix B]{GL}.

\section{$t'\neq0$: Approximate solution\label{sec:General-case}}

We were able to obtain only a formal exact solution for the PDF of
$C_{TA}(t',T')$ for $t'\neq0$ (see Appendix \ref{app:Formal-solution}).
Therefore, we resort to approximations. To start our analysis we divide
the integration interval $[0,T]$ into sojourn times $\tau_{j}$.
For convenience we redefine the first $t_{j}>T$ to be equal to $T$,
and denote its index by \emph{n}: $t_{n}\equiv T$. Accordingly, $\tau_{n}$
is redefined to be $T-t_{n-1}$ {[}cf. Fig. (\ref{cap:schematic}){]}.
Thus, for $i\leq n$ we write\begin{equation}
C_{TA}(t',T')=\frac{\sum_{i\,\,\mbox{odd}}^{n}\int_{t_{i-1}}^{t_{i}}I(t+t')\textrm{d}t}{T},\label{eq:Cta-intervals}\end{equation}
 where we used the initial condition that $I(t)=1$ at time $t=0$.
Hence $I(t)=1$ in $t_{i-1}<t<t_{i}$ when $i$ is odd, otherwise
it is zero. The summation in Eq. (\ref{eq:Cta-intervals}) is over
odd $i$'s, and $t_{n}=T$, namely $n-1$ in Eq. (\ref{eq:Cta-intervals})
is the random number of transitions in the interval $[0,T]$. From
Eq. (\ref{eq:Cta-intervals}) we see that the time averaged correlation
function, multiplied by \emph{T}, is a sum of the random variables
\begin{equation}
\int_{t_{i-1}}^{t_{i}}I(t)I(t+t')\textrm{d}t=\left\{ \begin{array}{ll}
\tau_{i}-t'+\mathcal{I}_{[t_{i},t_{i}+t']}t' & i\,\,\mbox{odd,}\,\,\tau_{i}>t'\\
\mathcal{I}_{[t_{i-1}+t',t_{i}+t']}\tau_{i} & i\,\,\mbox{odd,}\,\,\tau_{i}<t'\\
0 & i\,\,\mbox{even}.\end{array}\right.\label{eq:Int-i-th}\end{equation}
 Using Eqs. (\ref{eq:Cta-intervals}, \ref{eq:Int-i-th}) we find
an exact expression for the correlation function\begin{equation}
\begin{array}{c}
TC_{TA}\left(t',T'\right)={\displaystyle \sum_{i\,\,\mbox{odd}}^{n}}\tau_{i}-{\displaystyle \sum_{\begin{array}{c}
i\,\,\mbox{odd}\\
\tau_{i}<t'\end{array}}^{n}}(1-\mathcal{I}_{[t_{i-1}+t',t_{i}+t']})\tau_{i}\\
-\,\,\, t'{\displaystyle \sum_{\begin{array}{c}
i\,\,\mbox{odd}\\
\tau_{i}>t'\end{array}}^{n}}\left(1-\mathcal{I}_{[t_{i},t_{i}+t']}\right).\end{array}\label{eq:TC-gen}\end{equation}
The first term on the right hand side of this equation is $T^{+}$
the total time spent in state \emph{on} in the time interval $[0,T]$,
in the remaining two terms we have considered sojourn times $\tau_{i}$
larger or smaller than $t'$ separately.

The core idea of our approximate solution is to replace the time-averaged
intensities entering Eq. (\ref{eq:TC-gen}) by their mean-field value,
\emph{specific for a given realization}. Then for short $t'$ we replace
$\mathcal{I}_{[t_{i-1}+t',t_{i}+t']}$ and $\mathcal{I}_{[t_{i},t_{i}+t']}$
by $\mathcal{I}_{[0,T]}$, while for long $t'$ we use $\mathcal{I}_{[t',T']}$
instead. Some alternative approximations are given in Appendix \ref{app:Particular-solutions}.
In the following, we treat short and long $t'$ separately.

~

\subsection{Small $t'$}

Within the mean field theory, Eq. (\ref{eq:TC-gen}) is approximated
by\begin{equation}
TC_{TA}(t',T')=\mathcal{I}_{[0,T]}T-\left(1-\mathcal{I}_{[0,T]}\right)\left(t'N^{+}+\Sigma^{+}\right)\label{eq09}\end{equation}
 where $N^{+}$ is the number of odd (i.e. \emph{on}) intervals satisfying
$\tau_{i}\geq t'$ and $i\leq n$, while $\Sigma^{+}\equiv\sum_{i\,\,\mbox{odd},\tau_{i}<t'}^{n}\tau_{i}$
is the sum of all odd $\tau_{i}<t'$ and $i\leq n$. For any particular
realization $N^{+}$ will decrease with $t'$ in a step-wise fashion,
while $\Sigma^{+}$ will increase in a step-wise fashion. The term
$t'N^{+}+\Sigma^{+}$ in Eq. (\ref{eq09}), however, will be continuous.

We proceed by replacing $N^{+}$ and $\Sigma^{+}$ with their scaling
forms. $N^{+}$ should scale as $\sim n^{+}\int_{t'}^{T^{+}}\psi(\tau)d\tau$
and $\Sigma^{+}\sim n^{+}\int_{0}^{t'}\tau\psi(\tau)d\tau$, where
$n^{+}$ is the number of \emph{on} intervals comprising a given $T^{+}$.
First note that for $t'>T^{+}$, $N^{+}=0$ and $\Sigma^{+}=T^{+}$.
Second, we assume \begin{equation}
n^{+}\sim\frac{\sin\pi\theta}{\pi\theta}(T^{+})^{\theta}\label{eq:nplus-assume}\end{equation}
in analogy to the scaling of \emph{n} with \emph{T} (e.g., \cite{GL}).
Therefore, using Eq. (\ref{eq:power-law-decay}) we propose that for
$1\ll t'\leq T^{+}$\begin{equation}
N^{+}\approx\frac{\sin\pi\theta}{\pi\theta}\left[\left(\frac{T^{+}}{t'}\right)^{\theta}-1\right],\label{eq:Nplus}\end{equation}
and similarly,\begin{equation}
\Sigma^{+}\approx t'\left(\frac{T^{+}}{t'}\right)^{\theta}.\label{eq:Splus}\end{equation}

Finally, plugging Eqs. (\ref{eq:Nplus}, \ref{eq:Splus}) into Eq.
(\ref{eq09}) results in\begin{widetext}
\begin{equation}
C_{TA}(t',T')\simeq\left\{
\begin{array}{l l}
 {\mathcal{I}}_{[0,T]} \left\{ 1  - \left(1 - {\mathcal{I}}_{[0,T]} \right)
\left[ \left( {r \over {(1-r)\mathcal{I}}_{[0,T]} } \right)^{1 - \theta} \left( { \sin \pi \theta \over \pi \theta} + 1 \right) - { \sin \pi \theta \over \pi \theta}{ r \over {(1-r)\mathcal{I}}_{[0,T]} } \right] \right\} 
&\ t' < T^{+} \\
{\mathcal{I}}_{[0,T]} ^2 &\ t' > T^{+}.
\end{array}
\right.
\label{eq12}
\end{equation}
\end{widetext} Eq. (\ref{eq12}) yields the correlation function, however unlike
standard ergodic theories the correlation function here is a random
function since it depends on $\mathcal{I}_{[0,T]}$. 

The PDF of $C_{TA}(t',T')=z$ is now easy to find from the Lamperti
PDF of $\mathcal{I}_{[0,T]}=x$. Using the chain rule, and Eqs. (\ref{eq:Cta-TplusoverT},\ref{eq:lamperti},
\ref{eq12}):\begin{equation}
P_{C_{TA}(t',T')}(z(x))\approx\frac{\ell_{\theta}(x)}{\left|{\displaystyle \frac{dz(x)}{dx}}\right|}\label{eq:PDF-Cst-t'<<T}\end{equation}
which is a parametric representation of $P_{C_{TA}(t',T')}(z)$ ($dz/dx=dC_{TA}(t',T')/d\mathcal{I}_{[0,T]}$
is found from Eq. (\ref{eq12})).

In Figs. \ref{cap:theta0.3}, \ref{cap:theta0.5} and \ref{cap:theta0.8}
we plot the PDF of $C_{TA}(t',T')$ (dashed curves) together with
numerical simulations (diamonds) and find excellent agreement between
theory and simulation, for the cases where our approximations are
expected to hold $r<1/2$. In the above treatment we approximated
$\mathcal{I}_{[t',T']}$ by $\mathcal{I}_{[0,T]}$, which is legitimate
only for small enough $t'<T$, leading to a deterministic dependence
of $C_{TA}(t',T')$ on $\mathcal{I}_{[0,T]}$.

\textbf{Remark 1:} Note that in the ergodic case (in which we can
insert $\theta=1$ in the scaling relations) it follows that $C_{TA}(0,T')=\mathcal{I}_{[0,T]}=1/2$
for $r=0$ and $C_{TA}(0,T')=\mathcal{I}_{[0,T]}^{2}=1/4$ for any
$r\neq0$. This behavior reflects complete decorrelation of $I(t)$
and $I(t+t')$ for any (large enough) $t'$, irrespective of the value
of $T'$, as is indeed the case.

\textbf{Remark 2:} There is a certain similarity between Eq. (\ref{eq12})
and Eq. (\ref{eq:Cta-ens-av}) for small \emph{r}. Only qualitatively,
a realization with a given $\mathcal{I}_{[0,T]}$ can be viewed as
generated using $\psi_{\pm}(\tau)$ with $A_{+}\neq A_{-}$ (cf. Eq.
(\ref{eq:psi+-})), such that $P_{+}=\mathcal{I}_{[0,T]}$. See additional
discussion of Eq. (\ref{eq12}) in Appendix \ref{app:Notes-about-Eq.}.

\subsection{Large $t'$}

To understand the behavior of the PDF of the correlation function
for the limiting case $t'\approx T'\gg T$ the concept of persistence
is important (see Eq. (\ref{eq:p0-def})). Recall that the probability
of $I(t+t')=const$ on the interval $[t',T']$ grows to unity as $t'/T'\rightarrow1$.
Moreover, there is virtually no dependence on the signal values on
$t\in[0,T]$ and thus \begin{equation}
P_{C_{TA}(t',T')}(z)\approx\frac{1}{2}\ell_{\theta}(z)+\frac{1}{2}\delta(z-0).\label{eq:half-lamperti}\end{equation}
There is a collapse of half of the trajectories to a $\delta$-peak
at $z=0$, because of zero intensity of the signal on $[t',T']$ in
one of the two states, with probability $\rightarrow1/2$. In the
second case the signal will be unity throughout the interval $[t',T']$,
with probability $\rightarrow1/2$, while its relative \emph{on} time
distribution in $[0,T]$ is given by Lamperti PDF.

More generally, for $t'$ not so large, but still $t'>T$ we use the
mean-field, or decoupling approximation yielding from Eq. (\ref{eq:TC-gen})\begin{equation}
C_{TA}(t',T')\approx\mathcal{I}_{[0,T]}\mathcal{I}_{[t',T']}.\label{eq:Cta-big-t'}\end{equation}
To calculate the PDF of $C_{TA}(t',T')$ in Eq. (\ref{eq:Cta-big-t'})
we use two steps: (i) calculate the PDF of $\mathcal{I}_{[t',T']}=z$
which statistically depends on $\mathcal{I}_{[0,T]}$ (it is denoted
as $P_{\mathcal{I}_{[t',T']}}(z|\mathcal{I}_{[0,T]})$) and then (ii)
using the distribution of $\mathcal{I}_{[0,T]}$, which is the Lamperti's
PDF Eq. (\ref{eq:lamperti}), calculate the PDF of $C_{TA}(t',T')=z$: 

\begin{equation}
P_{C_{TA}(t',T')}(z)\sim\int_{0}^{1}\ell_{\theta}(x)P_{\mathcal{I}_{[t',T']}}\left(\left.\frac{z}{x}\right|x\right)\frac{dx}{x}.\label{eq:PDF-Cst}\end{equation}

Using the persistence probability Eq. (\ref{eq:p0-def}), we approximate
the conditional PDF of $\mathcal{I}_{[t',T']}=z$ for a given $\mathcal{I}_{[0,T]}$
in the case $T\ll t'$ by\\
\begin{widetext}

\begin{equation}
P_{\mathcal{I}_{[t',T']}}(z|\mathcal{I}_{[0,T]})\simeq\left[1-p_{0}\left(T,T'\right)\right]Q_{\mathcal{I}_{[t',T']}}\left(z\right)+p_{0}\left(T,T'\right)\left[\mathcal{I}_{[0,T]}\delta\left(z-1\right)+\left(1-\mathcal{I}_{[0,T]}\right)\delta\left(z\right)\right],\label{eq15}\end{equation}
 where $Q_{\mathcal{I}_{[t',T']}}\left(z\right)$ is the PDF of $\mathcal{I}_{[t',T']}$
conditioned that at least one transition occurs in $[T,T']$. In Eq.
(\ref{eq15}) we introduced the correlation between $\mathcal{I}_{[t',T']}$
and $\mathcal{I}_{[0,T]}$ through the dependence of the right hand
side of the equation on $\mathcal{I}_{[0,T]}$. We assumed that in
the case of no transitions in the time interval $[T,T']$, the probability
of the interval $[t',T']$ to be all the time either $on$ or $off$
(the only possible choices) is linearly proportional to the value
of $\mathcal{I}_{[0,T]}$.

The persistence probability controls also the behavior of \begin{equation}
Q_{\mathcal{I}_{[t',T']}}\left(z\right)\simeq\left[1-p_{0}\left(t',T'\right)\right]\Theta\left(0<z<1\right)+p_{0}\left(t',T'\right)\frac{\delta\left(z\right)+\delta\left(z-1\right)}{2}.\label{eq16}\end{equation}
 We assumed that if a transition occurs in the interval $[t',T']$
the distribution of $\mathcal{I}_{[t',T']}$ is uniform {[}i.e., $\Theta\left(0<z<1\right)=1$
if the condition in the parenthesis is correct{]}. This is a crude
approximation which is, however, reasonable for our purposes (however
when $\theta$ approaches 1, this approximation does not work). The
delta functions in Eq. (\ref{eq16}) arise from two types of trajectories:
If no transition occurs either $\mathcal{I}_{[t',T']}=1$ (state $on$)
or $\mathcal{I}_{[t',T']}=0$ (state $off$) with equal probability.
An asymptotically exact expression for $Q_{\mathcal{I}_{[t',T']}}\left(z\right)$
is given by Eq. (\ref{eq:Q-exact}) in Appendix \ref{app:Tplus-t1t2};
given the approximate nature of our derivations, however, we chose
to use Eq. (\ref{eq16}) because it is much simpler.

Finally, from Eqs. (\ref{eq15},\ref{eq16},\ref{eq:PDF-Cst}), and
using $\delta(a/x)=x\delta(a)$ for $x>0$, we obtain after some algebra
\begin{equation}
\begin{array}{l l}
 P_{ C_{TA}(t',T') } \left( z \right)  \simeq 
\left[ 1 - p_0\left(T, T' \right) \right] 
\left\{ \left[ 1 - p_0\left( t' , T' \right) \right] \int_z^1 { l_{\theta} \left( x \right) \over x} {\textrm{d}} x + 
{ p_0\left( t',T' \right) \over 2} \left[ l_{\theta} \left( z \right) + \delta\left( z \right) \right] \right\} 
+ p_0\left( T, T' \right) \left[ z l_{\theta}\left( z \right) + { \delta\left( z \right) \over 2} \right].
\end{array}
\label{eq17}
\end{equation}
\end{widetext} Note that to derive Eq. (\ref{eq17}) we used the fact that $\mathcal{I}_{[0,T]}$
and $\mathcal{I}_{[t',T']}$ are correlated. In Figs. \ref{cap:theta0.3},\ref{cap:theta0.5}
and \ref{cap:theta0.8} we plot these PDFs of $C_{TA}(t',T')$ (solid
curves) together with numerical simulations (diamonds) and find good
agreement between theory and simulation, for the cases where these
approximations are expected to hold, $r>1/2$. Eq. (\ref{eq:half-lamperti})
is recovered from Eq. (\ref{eq17}) in the limit of $t'/T'\rightarrow1$.

\section{Numerical simulations and comparison to approximations\label{sec:Numerical-simulations}}

We performed Monte Carlo simulations to generate distributions of
the time averaged correlation function $C_{TA}(t',T')$ for different
values of $r=t'/T'$ and with different $\theta$. Specifically, for
each chosen $\theta$ the function \[
\psi(\tau)=\left[\begin{array}{cc}
\theta\tau^{-1-\theta}, & \tau\geq1\\
\\0, & \tau<1\end{array}\right.\]
was used to generate random sojourn times until certain cumulative
time $T'\gg1$. This constitutes a single realization. Tens of thousands
of realizations were generated for each $\theta$. 

For each realization, $C_{TA}(t',T')$ was calculated for different
$t'$ using Eqs. (\ref{eq:Cst-def}) and (\ref{eq:sumsum-lmn}). To
check whether the PDF of $C_{TA}(t',T')$ depends only on \emph{r}
we used different $T'$. We also used the one-sided L$\grave{\textrm{e}}$vy
PDF for $\psi(\tau)$ and found that our results do not depend on
details of $\psi(\tau)$ besides the exponent $\theta$ of course.
In addition, we calculated $\left\langle C_{TA}(t',T')\right\rangle $
from our simulations and compared it to the theoretical result Eq.
(\ref{eq:meanCst-theor}). The agreement is excellent, as long as
$T=T'-t'\gg1$.

Some simulations are shown on Figures \ref{cap:theta0.3}, \ref{cap:theta0.5}
and \ref{cap:theta0.8} together with various theoretical approximations,
for $\theta=0.3,0.5\textrm{ and }0.8$ respectively. $\left\langle N\right\rangle $
is the number of transitions made until time $T'$, averaged over
realizations. Diamonds are simulated data. Solid lines for $r=0$
are $\ell_{\theta}(z)$ where $0\leq z\leq1$ are possible values
of $C_{TA}(0,T')$. Dashed and solid lines for $r\neq0$ are Eqs.
(\ref{eq:PDF-Cst-t'<<T}) and (\ref{eq17}) for $r\leq0.5$ and $r\geq0.5$,
respectively.

The discontinuity of the dashed lines, which can be noticed at small
values of $C_{TA}(t',T')$ for $r=0.1$ is due to the discontinuity
of the derivative in Eq. (\ref{eq:PDF-Cst-t'<<T}) at $\mathcal{I}_{[0,T]}=r/(1-r)$,
when $C_{TA}(t',T')$ becomes equal to $\mathcal{I}_{[0,T]}^{2}=r^{2}/(1-r)^{2}$,
which is very small for small \emph{r}. Overall, however, Eq. (\ref{eq:PDF-Cst-t'<<T})
agrees with the shown simulations for $r<0.5$.

Approximation (\ref{eq17}) works well for all $\theta$ values and
$r>0.5$, for which it was designed, and it can be seen that as \emph{r}
grows toward 1, the asymptotic result Eq. (\ref{eq:half-lamperti})
is approached. The assumption of uniform distribution of $\mathcal{I}_{[t',T']}$
for values between 0 and 1, used in Eq. (\ref{eq16}), is an oversimplification
when $\theta=0.8$, which is partly responsible for slight discrepancies
with the simulated data. Qualitatively, the PDF of $\mathcal{I}_{[t_{1},t_{2}]}$
is similar to $\ell_{\theta}(z)$ which starts growing a maximum at
$z=0.5$ for approximately $\theta>0.6$ and so Eq. (\ref{eq16})
is not very accurate for $\theta=0.8$. Also, here $T'\approx6\times10^{5}$
is not very large and therefore the simulated distributions haven't
completely reached their asymptotic forms (e.g., observe slight shape
differences between simulated data and theory for $r=0$).

Dot-dashed lines in Fig. \ref{cap:theta0.3} for $\theta=0.3$ are
based on Eq. (\ref{eq:Cst-small-theta}). They are shown only for
$r<0.5$; this approximation works well in the limit of small $\theta$
and \emph{r}. 

Our simulations show that for $\theta=0.5$ the PDF of $C_{TA}(t',T')=z$
is closely approximated by $2\left\langle C_{TA}(t',T')\right\rangle \ell_{0.5}(z)+\left(1-2\left\langle C_{TA}(t',T')\right\rangle \right)\delta(z)$.
Dotted lines in Fig. \ref{cap:theta0.5} are the nonsingular part
of this expression, i.e., Lamperti distributions normalized by the
relative mean. They are in good agreement with the data, and therefore
are hardly visible. We have no explanation for this fact, besides
the qualitative argument that as \emph{r} grows from zero, for lower
$\theta$ the left side of the distribution drops (cf. Fig. \ref{cap:theta0.3}),
while for higher $\theta$ it rises (cf. Fig. \ref{cap:theta0.8}),
and so somewhere between $\theta=0.3$ and $\theta=0.8$ it might
remain unchanged. For $t'/T'\rightarrow1$ this expression approaches
Eq. (\ref{eq:half-lamperti}).

For $r=0$ the PDF of $C_{TA}(t',T')$ is the Lamperti distribution.
As can be observed from comparison of the PDFs with $r=0$ and $r>0$
in Figs. \ref{cap:theta0.3} and \ref{cap:theta0.8}, the PDF of $C_{TA}(t',T')$
is shifted to the left as \emph{r} increases from zero. This is so
because small $\mathcal{I}_{[0,T]}$ values mean small proportion
of time spent \emph{on}, and there is a large probability that a small
$t'$ will yield zero correlation in such realizations. This is in
agreement with Eq. (\ref{eq:Cst-small-theta}). For larger $\theta$
values, there are more short and less long intervals covering the
time of ~{}``experiment'' $T'$ (as can be seen from Eq. (\ref{eq:short-prop}),
because $r^{1-\theta}$ increases toward 1 with growing $\theta$,
for any fixed $r>0$). Therefore, relatively small shift $t'$ (small
\emph{r}) will cause no significant effect in the case of small $\theta$,
dominated by large intervals, while in the case of large $\theta$
this small shift $t'$ will decorrelate many intervals, thus significantly
reducing the correlation function. Realizations with small $\mathcal{I}_{[0,T]}$
also will lose correlation faster for the same reason, leading to
a non-uniform visible deformation of the shape of the PDF of $C_{TA}(t',T')$.
Of course, as \emph{r} grows this simple picture breaks. However,
for \emph{r} approaching unity we recover another simple asymptotic
result (\ref{eq:half-lamperti}).

\subsection{2D histograms}

Two dimensional histograms, showing the frequency of events $C_{TA}(t',T')$
for a particular value of $\mathcal{I}_{[0,T']}$ are now considered.
These histograms show the correlation between $C_{TA}(t',T')$ and
$\mathcal{I}_{[0,T']}$. As we explained already for $r=0$ we have
$C_{TA}(t',T')=\mathcal{I}_{[0,T']}$, hence we have total correlation
in this simple case. When \emph{r} is small, our approximate solution
Eq. (\ref{eq12}) suggests a strong correlation between $C_{TA}(t',T')$
and $\mathcal{I}_{[0,T']}$. However, the arguments we used to derive
Eq. (\ref{eq12}) neglect fluctuations since they are based on our
non-ergodic mean field approximation. To check our mean field, and
to understand its limitations, the two dimensional histograms we consider
in this section are very useful. In addition, for large \emph{r} we
see from Eq. (\ref{eq17}), that according to the decoupling approximation,
the correlation between $C_{TA}(t',T')$ and $\mathcal{I}_{[0,T']}$
is expected to be weak, as is demonstrated indeed by correlation plots
in Fig. %
\begin{figure*}
\includegraphics[%
  scale=0.97]{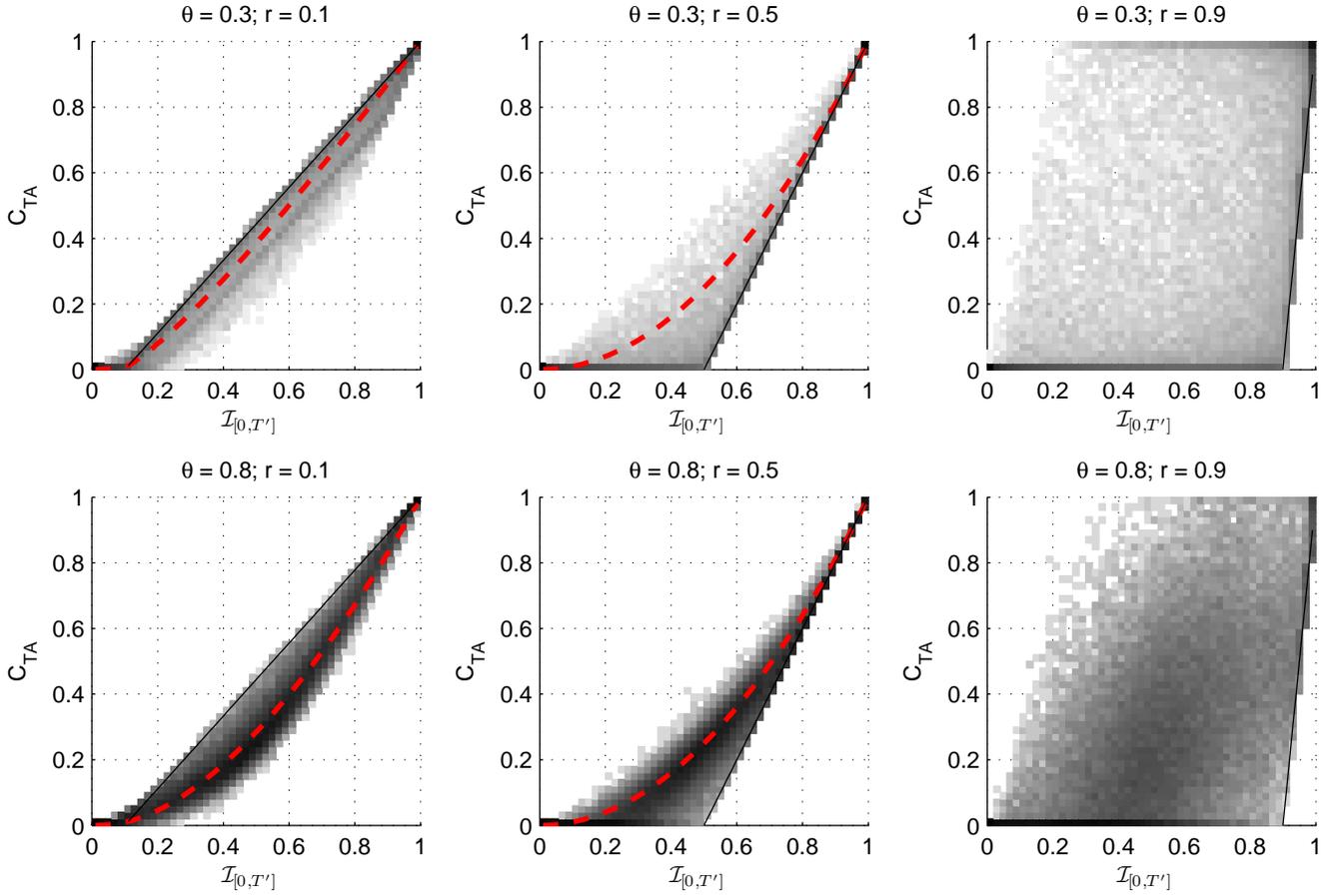}

\caption{\label{cap:Cta-2D}Distribution of $C_{TA}(t',T')$ as a function
of $\mathcal{I}_{[0,T']}$ for different values of \emph{r} and $\theta$.
The gray scale is changed logarithmically with the number of occurrences
inside a square bin. Darker regions mean higher occurrences. Dashed
lines are Eq. (\ref{eq12}) with $\mathcal{I}_{[0,T']}$ used instead
of $\mathcal{I}_{[0,T]}$. Full lines $C_{TA}(t',T')=\left(\mathcal{I}_{[0,T']}-r\right)/(1-r)$
are shown as well.}
\end{figure*}
\ref{cap:Cta-2D}. 

Leaving the details to Appendix \ref{app:Boundaries-of-Cta}, we can
derive the following rigorous boundaries (i.e., the inf and the sup)
of $C_{TA}(t',T')$: for $r\geq1/2$

\begin{equation}
\max\left\{ 0,\frac{\mathcal{I}_{[0,T']}-r}{1-r}\right\} \leq C_{TA}(t',T')\leq\min\left\{ 1,\frac{\mathcal{I}_{[0,T']}}{2(1-r)}\right\} ,\label{eq:Cta-bounds-r>1/2}\end{equation}
which is in agreement with Fig. \ref{cap:Cta-2D}. For $1/3\leq r<1/2$
we obtain\begin{equation}
\begin{array}{c}
\max\left\{ 0,\frac{\mathcal{I}_{[0,T']}+r-1}{1-r},\frac{2\mathcal{I}_{[0,T']}-r-1}{1-r}\right\} \leq C_{TA}(t',T')\\
\leq\min\left\{ \frac{2\mathcal{I}_{[0,T']}}{3(1-r)},\frac{\mathcal{I}_{[0,T']}+1-2r}{2(1-r)}\right\} .\end{array}\label{eq:Cta-bounds-1/3<r<1/2}\end{equation}

For small \emph{r}, on one hand the matters become more complicated,
so our argument is more qualitative . First notice that if $T_{[0,T']}^{-}\equiv T'-T_{[0,T']}^{+}$
is small enough then all the \emph{off} intervals can lie inside $[t',T]$
and be used twice (once in $I(t)$ and once in $I(t+t')$, for $r\neq0$)
to multiply \emph{on} intervals, hence $C_{TA}(t',T')\geq\left(T-2T_{[0,T']}^{-}\right)/T=\left(2\mathcal{I}_{[0,T']}-r-1\right)/(1-r)$.
In most cases, small $T_{[0,T']}^{-}$ means that the last sojourn
interval (going up to time $T'$) is in state \emph{on} and all the
\emph{off} intervals are inside $[0,T]$, so that $C_{TA}(t',T')\leq\left(T-T_{[0,T']}^{-}\right)/T=\left(\mathcal{I}_{[0,T']}-r\right)/(1-r)$.
Compare this to Eq. (\ref{eq:Cst-small-theta}) derived in Appendix
\ref{app:Particular-solutions}. It is argued there that this value
of $C_{TA}(t',T')$ will be achieved more often for lower $\theta$,
in agreement with Fig. \ref{cap:Cta-2D}. This is \emph{not} a rigorous
upper bound, though; see Appendix \ref{app:Boundaries-of-Cta}. If
$T_{[0,T']}^{+}$ is small enough then the lower bound will be zero.
The sufficient (but not necessary, in general) condition to achieve
zero is $T_{[0,T']}^{+}\leq1/2$ as then we can construct a trajectory
by choosing zero intensity at the time $t+t'$ if it is 1 at time
\emph{t}, and vice versa. 

On the other hand, for very small \emph{r} (but $t'$ can be large)
we know that $C_{TA}(t',T')$ is almost unchanged, as the whole signal
is dominated by relatively few largest sojourn intervals (cf. Eq.
(\ref{eq:short-prop})); hence $C_{TA}(t',T')$ will be close to $\mathcal{I}_{[0,T']}$.
The rigorous bounds are, therefore, hardly reached.

\textbf{Remark:} Our approximation Eq. (\ref{eq12}), with $\mathcal{I}_{[0,T]}$
replaced by $\mathcal{I}_{[0,T']}$, is shown by the dashed lines
on Fig. \ref{cap:Cta-2D}. For $r=1/2$ it actually reduces to $C_{TA}(t',T')=\mathcal{I}_{[0,T']}^{2}$,
which works better for higher $\theta$, when the non-ergodicity is
weaker. In fact, a more precise way to find $C_{TA}(t',T')$ in this
case is using Eq. (\ref{eq:Cta-big-t'}), but then there is no simple
formula connecting $C_{TA}(t',T')$ and $\mathcal{I}_{[0,T]}$ like
Eq. (\ref{eq12}).

\subsection{Power spectrum}

\begin{figure}
\includegraphics[%
  width=1.0\columnwidth,
  keepaspectratio]{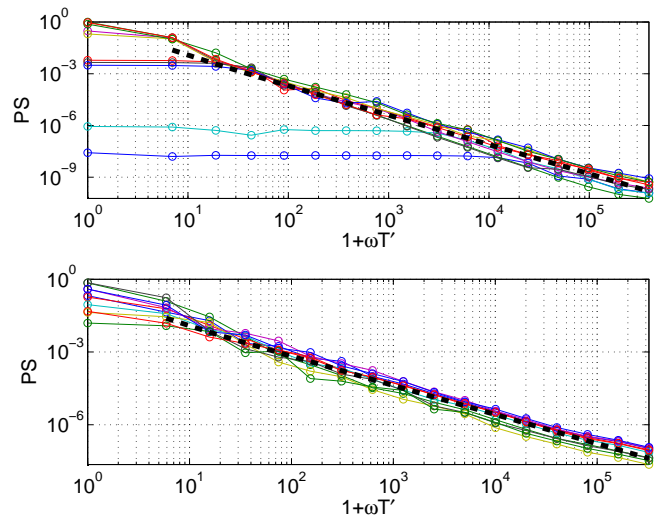}

\caption{\label{cap:PSD-realiz}Power spectrum $S(\omega)/T'$ for ten typical
realizations shown in Fig. \ref{cap:Cta-realiz}, for $\theta=0.3$
(top) and $\theta=0.8$ (bottom). Data for each realization are averaged
in exponentially increasing with $\omega$ bins. Each curve is normalized
in such a way that at $\omega=0$ the PS equals $\mathcal{I}_{[0,T']}^{2}$.
The PS is random due to non-ergodicity of underlying process. Dashed
lines are given by Eq. (\ref{eq:PSDens}) and scale as $\omega^{\theta-2}$.
The abscissas are $1+\omega T'$ in order to show the value of PS
at zero frequency on a log-log plot.}
\end{figure}

It is useful to look at power spectra (PS) of generated intensity
signals \cite{Pelton04,Davidsen,Zumofen93,Schriefl05}. Power spectrum
is defined as\begin{equation}
S(\omega)=\frac{\tilde{I}(\omega)\tilde{I}(-\omega)}{T'},\label{eq:PS-def}\end{equation}
where $\tilde{I}(\omega)$ is Fourier transform of $I(t)$ (cf. Eq.
(\ref{eq:FI})). We calculate such PS and find, as expected, that
they too exhibit a nonergodic behavior, as shown in Fig. \ref{cap:PSD-realiz}.
Each PS is random and does not fall on the ensemble averaged curve
(dashed line) even after averaging the data in large frequency windows.
Note that for smaller $\theta$ the PS values for a given $\omega$
are spread wider, which is a reflection of a wider distributions of
correlation functions (cf. Figs. \ref{cap:theta0.3}, \ref{cap:theta0.8}).
In light of the scaling $C_{TA}(t',T')\sim A-Br^{1-\theta}$ in expressions
(\ref{eq:Cta-ens-av}) and (\ref{eq12}) for small enough \emph{r}
(but for $t'$ as large as desired, as long as $T'$ is large enough)
we can argue that the PS will scale as (cf. Eq. (\ref{eq:S-Cta}))\begin{equation}
\begin{array}{ccc}
S(\omega) & \sim & -2B(T')^{\theta-1}\textrm{Re}\int_{0}^{T'}(t')^{1-\theta}e^{-i\omega t'}dt'\qquad\qquad\quad\\
 & \approx & 2BT'\cos(\pi\theta/2)\Gamma(2-\theta)(\omega T')^{\theta-2}\propto\omega^{\theta-2}\end{array}\label{eq:PS-scaling}\end{equation}
as long as $\omega\gg1/T'$ (term \emph{A} in $C_{TA}(t',T')$ leads
to a term $2AT'\sin\omega T'/(\omega T')$ which is zero for all $\omega\neq0$
used in calculating discrete power spectrum). This is indeed the case,
as illustrated in Fig. \ref{cap:PSD-realiz}. In Eq. (\ref{eq:PS-scaling})
we estimated the PS by Fourier transforming the correlation function,
implying the well-known Wiener-Khintchine theorem. This theorem, however,
is assumed valid only for stationary processes and for ensemble averaged
correlation functions and spectra. Nevertheless, one can say that
with respect to short sojourn times each realization is identical,
and the observed non-ergodicity is due to necessarily poor statistics
of long intervals (leading, in particular, to different values of
\emph{B} for different realizations). See Appendix \ref{app:Formal-solution}
for discussion on a generalized Wiener-Khintchine theorem.%

For the ensemble-averaged spectrum, using the value $B=\sin(\pi\theta)/(4\pi\theta(1-\theta))$
from Eq. (\ref{eq:Cta-ens-av}), \begin{equation}
\frac{\left\langle S(\omega)\right\rangle }{T'}\sim\frac{\cos(\pi\theta/2)}{2\Gamma(1+\theta)}(\omega T')^{\theta-2},\,\,\,\omega T'\gg1.\label{eq:PSDens}\end{equation}
This line is also shown in Fig. \ref{cap:PSD-realiz}.

\section{Summary}

We investigated autocorrelation of a dichotomous random process governed
by identical waiting time distributions of its two states, characterized
by zero and nonzero intensity. We considered the case of a power law
waiting time with exponent $\theta$ lying between 0 and 1, as this
choice is of considerable practical interest. This process is a one-dimensional
L\'evy walk process. Such power law distributions are experimentally
observed, as discussed in the Introduction. These distributions lead
to aging and non-ergodicity and in particular, to a distribution of
possible values of a single trajectory two-time correlation function
for fixed times, even in the limit when these times go to infinity.
This is in striking contrast to the standard situation in which correlation
function asymptotically assumes only one possible value for fixed
times, equal to the ensemble average (ergodicity).

For our theoretical analysis of distributions of correlation functions
we used the non-ergodic mean-field and the decoupling approximation,
Eqs. (\ref{eq09}) and (\ref{eq:Cta-big-t'}), in which various temporal
averages of the intensity were replaced by the total time averaged
intensity $\mathcal{I}_{[0,T]}$ or $\mathcal{I}_{[t',T']}$, specific
for each realization. We then expressed the correlation function as
a (deterministic or random) function of this time average. This enabled
us to derive approximate results for the distributions of correlation
functions from known distributions of time averaged intensity. We
also related power spectra of single trajectories to the time averaged
correlation functions, and demonstrated their nonergodicity as well
as universal scaling which is a function of the exponent $\theta$
only. Our results agree well with numerical simulations, and, importantly,
clarify the nature of the investigated non-ergodicity. Generalizations
of our approach to situations with different \emph{on} and \emph{off}
time distributions are possible. 

In the context of blinking nanocrystals, we showed \cite{MB_JCP}
that the exponent $\theta=1/2$ is a result of a simple model of first
passage time of charge carrier in three dimensions, based on standard
diffusion. The experiments \cite{Dahan,Zumofen04,Xie,Weitz,Kuno}
show, that rather generally, power law sojourn times describe dynamics
of single particles in diverse systems. Since power law sojourn times
(not necessarily for a two state process) lead to non-ergodic behavior,
we expect that stochastic theories of ergodicity breaking will play
an increasingly important role in the analysis of single particle
experiments.

\begin{acknowledgments}
This work was supported by National Science Foundation award CHE-0344930.
EB also thanks Center for Complexity Science, Israel.
\end{acknowledgments}
\appendix

\section{Formal solution\label{app:Formal-solution}}

We express the numerator in Eq. (\ref{eq:Cst-def}) through the cumulative
renewal (transition) times $t_{j}$ by noting that $I(t)=const$ on
intervals $t_{n}<t<t_{n+1}$, while $I(t+t')=const$ on intervals
$t_{m}-t'<t<t_{m+1}-t'$ and keeping in mind the restriction $0<t<T$.
For convenience, we redefine the first $t_{j}$ which is $>T'$ to
be equal to $T'$. The index of this $t_{j}$ is denoted by $N$.
Note that $\tau_{N}\equiv T'-t_{N-1}$ is not distributed according
to $\psi(\tau)$. The temporal durations (lengths) of intervals where
both $I(t)$ and $I(t+t')$ are constant, are then\begin{equation}
l_{mn}(t')=\max\left\{ 0,\min(t_{n+1},t_{m+1}-t')-\max(t_{n},t_{m}-t')\right\} \label{eq:l-mn-def}\end{equation}
and in particular\[
l_{nn}=\max\left\{ 0,\tau_{n+1}-t'\right\} .\]
Obviously, $l_{m<n,n}=0$ and hence\begin{equation}
\int_{0}^{T'-t'}I(t)I(t+t')dt=\sum_{\begin{array}{c}
n=0,\\
n\,\textrm{even}\end{array}}^{N-1}\sum_{\begin{array}{c}
m=n,\\
m\,\textrm{even}\end{array}}^{N-1}l_{mn}.\label{eq:sumsum-lmn}\end{equation}
 Here, and throughout the article, we assume that the process starts
in state \emph{on}. This assumption is clearly asymptotically negligible,
and is made here simply for purposes of notation.

Using the cumulative PDF of $\{\tau_{1},...,\tau_{N-1}\}$ and $N-1$
under the constraint $t_{N-1}<T'$ \cite{GL} (and because $\textrm{Prob}[t_{N-1}=T']=0$)
we can write formally the PDF of $TC_{TA}(t',T')=x$ as\[
\begin{array}{c}
f_{TC_{TA}(t',T')}(x)={\displaystyle \sum_{N=1}^{\infty}}\left({\displaystyle \prod_{k=1}^{N-1}}{\displaystyle \int_{0}^{\infty}}\psi(\tau_{k})d\tau_{k}\right)\mathcal{P}(t_{N}=T')\times\\
\\\times\Theta(T'-t_{N-1})\delta\left(x-{\displaystyle \sum_{\begin{array}{c}
n=0,\\
n\,\textrm{even}\end{array}}^{N-1}\sum_{\begin{array}{c}
m=n,\\
m\,\textrm{even}\end{array}}^{N-1}}l_{mn}(t')\right),\end{array}\]
where $\Theta$ is Heaviside step function, $\mathcal{P}(t_{N}=T')$
is the probability that no transition occurred between $t_{N-1}$
and $T'$, and $\delta$ is the Dirac delta.

In order to get rid of the max and min functions in Eq. (\ref{eq:l-mn-def}),
one can perform Laplace transform of Eq. (\ref{eq:l-mn-def}) with
respect to $t'$ ($t'\rightarrow u$) and write similar expression
for the PDF of $T\hat{C}_{TA}(u,T')$ (for real \emph{u}). It can
be shown that\begin{equation}
\hat{l}_{m>n,n}(u)=\frac{e^{-u(t_{m+1}-t_{n})}(e^{u\tau_{m+1}}-1)(e^{u\tau_{n+1}}-1)}{u^{2}}\label{eq:Llmn}\end{equation}
and\begin{equation}
\hat{l}_{nn}(u)=\frac{u\tau_{n+1}-1+e^{-u\tau_{n+1}}}{u^{2}}.\label{eq:Llnn}\end{equation}
We could not, unfortunately, utilize these expressions to calculate
$f_{TC_{TA}(t',T')}(x)$ or $f_{T\hat{C}_{TA}(u,T')}(x)$ and therefore
have to resort to various approximations.

\subsection*{Relation to power spectrum}

We derive a generalized form of Wiener-Khinchine theorem for nonergodic
nonstationary processes. In analogy to the numerical spectral analysis
of a time series, we assume here that the intensity signal is identically
zero outside of the interval $[0,T']$. Then the Fourier transform
of intensity is defined as\begin{equation}
\tilde{I}(\omega)={\displaystyle \int_{-\infty}^{\infty}}I(t)e^{-i\omega t}dt={\displaystyle \int_{0}^{T'}}I(t)e^{-i\omega t}dt\label{eq:FI}\end{equation}
and the power spectrum of a realization is defined in Eq. (\ref{eq:PS-def}).
From Eqs. (\ref{eq:FI}) and (\ref{eq:PS-def})\[
T'S(\omega)={\displaystyle \int_{0}^{T'}}I(t_{1})e^{-i\omega t_{1}}dt_{1}{\displaystyle \int_{0}^{T'}}I(t_{2})e^{i\omega t_{2}}dt_{2}.\]
We now divide the integration over $t_{2}$ into two parts and replace
the order of integration in the first part:\[
\begin{array}{ccc}
{\displaystyle \int_{0}^{T'}}dt_{1}{\displaystyle \int_{0}^{T'}}dt_{2} & = & {\displaystyle \int_{0}^{T'}}dt_{1}{\displaystyle \int_{0}^{t_{1}}}dt_{2}+{\displaystyle \int_{0}^{T'}}dt_{1}{\displaystyle \int_{t_{1}}^{T'}}dt_{2}\\
\\ & = & {\displaystyle \int_{0}^{T'}}dt_{2}{\displaystyle \int_{t_{2}}^{T'}}dt_{1}+{\displaystyle \int_{0}^{T'}}dt_{1}{\displaystyle \int_{t_{1}}^{T'}}dt_{2}.\end{array}\]
Swapping names $t_{1}$ and $t_{2}$ in the first part thus yields\[
T'S(\omega)={\displaystyle \int_{0}^{T'}}dt_{1}{\displaystyle \int_{t_{1}}^{T'}}dt_{2}I(t_{1})I(t_{2})[e^{i\omega(t_{1}-t_{2})}+e^{i\omega(t_{2}-t_{1})}]\]
and introducing $t=t_{1}$ and $t'=t_{2}-t_{1}$ results in\[
S(\omega)=\frac{1}{T'}{\displaystyle \int_{0}^{T'}}dt{\displaystyle \int_{0}^{T'-t}}dt'I(t)I(t+t')[e^{-i\omega t'}+e^{i\omega t'}].\]
In a similar fashion, we can write the Laplace transform of \begin{equation}
K(t',T')\equiv(T'-t')C_{TA}(t',T')={\displaystyle \int_{0}^{T'-t'}}dtI(t)I(t+t')\label{eq:K}\end{equation}
with respect to $t'$ as\[
\begin{array}{ccc}
\hat{K}(u,T') & = & {\displaystyle \int_{0}^{\infty}}dt'e^{-ut'}{\displaystyle \int_{0}^{T'-t'}}dtI(t)I(t+t')\\
\\ & = & {\displaystyle \int_{0}^{T'}}dt{\displaystyle \int_{0}^{T'-t}}dt'I(t)I(t+t')e^{-ut'}\end{array}\]
and it becomes evident that \begin{equation}
T'S(\omega)=\hat{K}(i\omega,T')+\hat{K}(-i\omega,T').\label{eq:WKT}\end{equation}
This is a generalization of the Wiener-Khintchine theorem stating
that the power spectrum is given by cosine Fourier transform of a
correlation function. But while this theorem is used for ensemble-averaged
correlation functions of stationary processes, here we have a similar
relation for a non-stationary process, and without ensemble averaging.
Note that the dependence on $T'$ is preserved, in contrast to the
regular Wiener-Khintchine theorem.

From Eq. (\ref{eq:K}) it follows that \[
\hat{K}(u,T')=T'\hat{C}_{TA}(u,T')+\frac{\partial\hat{C}_{TA}(u,T')}{\partial u}\]
and for large $uT'\gg1$, $\hat{C}_{TA}(u,T')\approx\hat{K}(u,T')/T'$,
leading finally to \begin{equation}
\begin{array}{ccc}
S(\omega) & = & \hat{C}_{TA}(i\omega,T')+\hat{C}_{TA}(-i\omega,T')\\
\\ & + & {\displaystyle \frac{\partial[\hat{C}_{TA}(i\omega,T')-\hat{C}_{TA}(-i\omega,T')]}{i\partial\omega}}\\
\\ & \approx & \hat{C}_{TA}(i\omega,T')+\hat{C}_{TA}(-i\omega,T')\end{array}\label{eq:S-Cta}\end{equation}
for large $\omega T'\gg1$. Note that for a single trajectory correlation
defined as $\left(\int_{0}^{T'-t'}I(t)I(t+t')dt\right)/T'$ instead
of Eq. (\ref{eq:Cst-def}) the generalized Wiener-Khintchine relation
is exact for any $\omega$ (cf. Eq. (\ref{eq:WKT})).

As an illustration, consider now our case of the \emph{on}-\emph{off}
process. Fourier transform of an intensity $I(t)$ for a realization
is:\[
\tilde{I}(\omega)={\displaystyle \frac{1}{-i\omega}\sum_{\begin{array}{c}
n=0,\\
n\textrm{ even}\end{array}}^{N-1}}e^{-i\omega t_{n+1}}(1-e^{i\omega\tau_{n+1}}).\]
Then it is straightforward to show that \[
T'S(\omega)=\hat{K}(i\omega,T')+\hat{K}(-i\omega,T'),\]
where $\hat{K}(u,T')$ can be found utilizing Eqs. (\ref{eq:sumsum-lmn}),
(\ref{eq:Llmn}) and (\ref{eq:Llnn}). This is a particular case of
the general relation (\ref{eq:WKT}).

\section{Distribution of $T_{[a,b]}^{+}$\label{app:Tplus-t1t2}}

Here we present asymptotically exact formula for a distribution of
\emph{on} times on an arbitrary interval $[a,b]$, where $a$ and
$b-a$ are large enough. We denote the first renewal time after $a$
by $\nu$. We have to take two possibilities into account. First is
that there was at least one renewal inside the interval and then $a<\nu<b$.
Thus \[
T_{[a,b]}^{+}=Y+T_{[\nu,b]}^{+}\]
 where \emph{Y} is the \emph{on} time from $a$ till first renewal
$\nu$. Asymptotically, \emph{Y} is independent of initial conditions
and its PDF is\[
f_{Y}(y)=\frac{1}{2}\delta(y-(\nu-a))+\frac{1}{2}\delta(y-0),\]
where the two Dirac deltas correspond to being in state \emph{on}
or \emph{off}.

After renewal at time $\nu$, again asymptotically, we can use the
PDF of $T_{[\nu,b]}^{+}$ which is also independent of its {}``initial
condition'', i.e., the value of \emph{Y} being 0 or 1. Then the PDF
of $T_{[\nu,b]}^{+}/(b-\nu)$ is given by the Lamperti $\ell_{\theta}$
and therefore for any fixed $\nu$\begin{equation}
\begin{array}{c}
f_{T_{[a,b]}^{+}}(x|\nu;\nu<b)\\
\\=\frac{1}{2(b-\nu)}\left[\ell_{\theta}\left(\frac{x-(\nu-a)}{b-\nu}\right)+\ell_{\theta}\left(\frac{x}{b-\nu}\right)\right].\end{array}\label{eq:Tplust1<t<t2}\end{equation}

The second possibility is that $\nu>b$ and in this case, clearly,\[
f_{T_{[a,b]}^{+}}(x|\nu;\nu>b)=\frac{1}{2}\delta(x-(b-a))+\frac{1}{2}\delta(x-0).\]

Introducing the PDF of the forward recurrence time, $f_{E}(\nu-a;a)$,
which is the PDF of having to wait for the first renewal after time
$a$ for a period of time $\nu-a$ \cite{GL}, we finally obtain\[
\begin{array}{c}
f_{T_{[a,b]}^{+}}(x)={\displaystyle \int_{a}^{b}}f_{T_{[a,b]}^{+}}(x|\nu)f_{E}(\nu-a;a)d\nu\\
\\+\frac{1}{2}\left(\delta(x-(b-a))+\delta(x-0)\right){\displaystyle \int_{b}^{\infty}}f_{E}(\nu-a;a)d\nu,\end{array}\]
with $f_{T_{[a,b]}^{+}}(x|\nu)$ in the first integral given by Eq.
(\ref{eq:Tplust1<t<t2}). The last integral\begin{equation}
p_{0}(a,b)={\displaystyle \int_{b}^{\infty}}f_{E}(\nu-a;a)d\nu\label{eq:p0-int-fE}\end{equation}
defines the persistence probability $p_{0}(a,b)$, for $b\geq a$.
The function $f_{E}(\nu-a;a)$ is equal to $\psi(\nu)$ for $a=0$.
For large $a$, it is the Dynkin function \cite{GL,Feller,MB_JCP}\[
f_{E}(\nu-a;a)\sim\frac{\sin\pi\theta}{\pi}\frac{1}{\left(\frac{\nu}{a}-1\right)^{\theta}\nu},\]
 and then $p_{0}(a,b)$ can be written as in Eq. (\ref{eq:p0-def}). 

Finally, the PDF of $\mathcal{I}_{[a,b]}=z$ is\begin{equation}
Q_{\mathcal{I}_{[a,b]}}(z)=(b-a)f_{T_{[a,b]}^{+}}((b-a)z).\label{eq:Q-exact}\end{equation}

\section{Particular solutions\label{app:Particular-solutions}}

We consider here two situations which can be analyzed differently
from the approach presented in Section \ref{sec:General-case}. This
analysis helps understanding the structure of the correlation functions.

\subsection{Extremely small $t'$}

If $t'<\tau_{i},\,\, i=1,...,N$ then, using Appendix \ref{app:Formal-solution},
we obtain $l_{m>n+1,n}=0$, $l_{nn}=\tau_{n+1}-t'$ and\begin{equation}
TC_{TA}(t',T')=T_{[0,T']}^{+}-\left[\frac{N+1}{2}\right]_{int}t'\approx T_{[0,T']}^{+}-\frac{Nt'}{2}\label{eq:TC-small-t'}\end{equation}
where the subscript \emph{int} indicates that {[}...{]} denote integer
part. It is easy to see that in this case, $T_{[0,T']}^{+}>0$ for
$t'>0$ and, moreover, $0<TC_{TA}(t',T')\leq T$, as it should (because
$N\geq1$).

The fraction of time $T'$ covered by short intervals $\tau_{i}<t'$
scales as\begin{equation}
\frac{\int_{0}^{t'}t\psi(t)dt}{\int_{0}^{T'}t\psi(t)dt}\approx\left(\frac{t'}{T'}\right)^{1-\theta}=r^{1-\theta}\label{eq:short-prop}\end{equation}
for large enough $t'$ and $T'$. Hence the contribution of these
short intervals is negligible if $t'\ll T'$, although $t'$ is large
(in contrast to the case when the mean sojourn time is finite and
the fraction of time covered by intervals shorter than $t'$ grows
to 1 as $t'$ increases, irrespective of the ratio $t'/T'$). Therefore,
we argue that Eq. (\ref{eq:TC-small-t'}) can still be used when $t'\ll T'$,
if by $N$ we understand the number $N_{eff}$ of intervals longer
than $t'$. It is important, however, to distinguish these coarsened
intervals from the original intervals $\tau_{i}$. The durations $\tau_{eff}$
of the coarsened intervals are not governed by $\psi(\tau)$. For
a power law $\psi(\tau)$ we expect that their durations are still
governed by a power law PDF with the same exponent $\theta$. Nevertheless,
it is questionable to use asymptotic expressions for $N_{eff}$ as
a function of $t'/T'$, constructed in a fashion used in Section \ref{sec:General-case},
because the PDF of $\tau_{eff}$ does not have to be a power law for
$\tau_{eff}\sim t'$ (and it is zero for $\tau_{eff}<t'$).

\subsection{Small $\theta$ and intermediate $t'$}

It is possible to make an exact calculation if there exists an interval
number \emph{k}, and for $t'$ such that\[
\sum_{\begin{array}{c}
i=1\\
i\neq k\end{array}}^{N}\tau_{i}<t'<\tau_{k}.\]
Ubiquitous realization of this condition could be expected for small
$\theta$, when the longest interval often approaches the {}``experimental''
time $T'$. Then, for $m>n$ using\[
\begin{array}{ccc}
t_{n}-(t_{m}-t') & = & t'-{\displaystyle \sum_{i=n+1}^{m}}\tau_{i}\\
\\t_{n+1}-(t_{m+1}-t') & = & t'-{\displaystyle \sum_{i=n+2}^{m+1}}\tau_{i}\end{array}\]
yields\[
l_{m>n+1,n}=\delta_{n,k-1}\tau_{m+1}+\delta_{m,k-1}\tau_{n+1};\]
also\[
l_{nn}=\delta_{n,k-1}(\tau_{k}-t'),\]
where $\delta_{ij}$ is Kronecker delta, and hence\begin{equation}
C_{TA}(t',T')=\left[\begin{array}{cc}
0, & k\textrm{ even}\\
\\{\displaystyle \frac{\mathcal{I}_{[0,T']}-r}{1-r}}, & k\textrm{ odd.}\end{array}\right.\label{eq:Cst-small-theta}\end{equation}
In the case of odd \emph{k}, $T_{[0,T']}^{+}\geq\tau_{k}>t'$ so that
always $0\leq C_{TA}(t',T')\leq1$, as it should. This particular
solution also plays an important role in defining the boundaries of
the two-dimensional correlation plots discussed in Section \ref{sec:Numerical-simulations}.

\section{Notes about Eq. (\ref{eq12})\label{app:Notes-about-Eq.}}

In this appendix, we discuss some approximations involved in the derivation
of Eq. (\ref{eq12}) and some of its shortcomings.

We begin with Eq. (\ref{eq:nplus-assume}). Scaling behavior $n\propto T^{\theta}$,
where \emph{n} is the number of transitions up to time \emph{T}, is
well-known for $0<\theta<1$ (e.g., \cite{GL}). However, the distribution
of \emph{n} is wide and its standard deviation is also known to scale
as $T^{\theta}$. For our purposes, we want to represent this standard
deviation as arising from two contributions. First is that \emph{n}
depends on $T^{+}\equiv T_{[0,T]}^{+}$, while second contribution
is that for any fixed $T^{+}$ there still is a distribution of \emph{n}
values. We can approximate the first contribution by writing $n\propto\left(T^{+}(T-T^{+})/T\right)^{\theta}$.
Since $T^{+}\propto T$, this formula does not contradict standard
scaling $n\propto T^{\theta}$, and it is at least in qualitative
agreement with our numerical simulations. To justify it we observe
that when $T^{+}\ll T$ then there is probably a large interval of
state \emph{off}, which covers almost all the time \emph{T}. If we
remove this large interval then the remaining total time will be of
the order of $T^{+}$, while the number of intervals will essentially
remain unchanged (will decrease by 1). Hence, in this case $n\propto T^{+}$.
Similar arguments apply when $T^{-}\equiv T-T^{+}\ll T$, leading
to the proposed scaling. We neglect the second contribution, although
it is not small. In Eq. (\ref{eq:nplus-assume}) we used $n^{+}\propto\left(T^{+}\right)^{\theta}$,
while the scaling part $n^{+}\propto\left((T-T^{+})/T\right)^{\theta}$
was absorbed in the coefficients. We also should ideally recover the
relation \[
n\sim\frac{\sin\pi\theta}{\pi\theta}T^{\theta},\]
which leads to \begin{equation}
\frac{\sin\pi\theta}{\pi\theta}\sim2\int_{0}^{1}\frac{n^{+}(z)}{T^{\theta}}\ell_{\theta}(z)dz,\label{eq:nplus-integrate}\end{equation}
where $z\equiv T^{+}/T$ and the factor of 2 arises because $n^{+}\sim n/2$.
In case of Eq. (\ref{eq:nplus-assume}) we have $n^{+}\sim(Tz)^{\theta}\sin\pi\theta/(\pi\theta)$
and relation (\ref{eq:nplus-integrate}) is fulfilled approximately.
One can instead approximate $n^{+}\sim a(Tz)^{\theta}$ or maybe $n^{+}\sim b(1-z)^{\theta}(Tz)^{\theta}$
where \emph{a} or \emph{b} will be determined from Eq. (\ref{eq:nplus-integrate}).
Alternatively, \emph{a} or \emph{b} can be determined by equating
the ensemble average of Eq. (\ref{eq12}) for small \emph{r} with
Eq. (\ref{eq:Cta-ens-av}) for $P_{+}=1/2$.

There are two noticeable shortcomings of Eq. (\ref{eq12}). One of
them regarding the discontinuity of its derivative is mentioned in
Section \ref{sec:Numerical-simulations}. The other one becomes clear
if one considers the complementary intensity signal $J(t)=1-I(t)$.
It follows from Eq. (\ref{eq:Cst-def}) that\begin{equation}
C_{I}(t',T')=\mathcal{I}_{[0,T]}+\mathcal{I}_{[t',T']}-1+C_{J}(t',T'),\label{eq:CiCj}\end{equation}
where $C_{I}(t',T')\equiv C_{TA}(t',T')$ and $C_{J}(t',T')$ is the
time-averaged correlation of signal $J(t)$. Eq. (\ref{eq12}) is
written for $C_{I}(t',T')$, but analogous equation can be written
for $C_{J}(t',T')$ as well, where $\mathcal{I}_{[0,T]}$ is replaced
by $1-\mathcal{I}_{[0,T]}$. Then, unfortunately, the relation (\ref{eq:CiCj})
will not hold in general. It will be satisfied trivially if $t'=0$,
or if $t'$ is large enough so that one can use the second line of
Eq. (\ref{eq12}) for both $C_{I}(t',T')$ and $C_{J}(t',T')$ (more
precisely, if $C_{I}(t',T')\sim\mathcal{I}_{[0,T]}\mathcal{I}_{[t',T']}$
as in Eq. (\ref{eq:Cta-big-t'}) and also $C_{J}(t',T')\sim(1-\mathcal{I}_{[0,T]})(1-\mathcal{I}_{[t',T']})$).

\section{Boundaries of $C_{TA}(t',T')$\label{app:Boundaries-of-Cta}}

Let us first consider the simpler case of $r\geq1/2$: then $T\geq t'$.
If $T_{[0,T']}^{-}\equiv T'-T_{[0,T']}^{+}\leq t'-T$, or equivalently
$\mathcal{I}_{[0,T']}\geq2(1-r)$ then all the \emph{off} intervals
can be placed inside the interval $[T,t']$ and hence $C_{TA}(t',T')$
can attain its maximal value of 1, which we will write as $C_{TA}(t',T')\leq1$,
meaning that the limit is achievable. For $\mathcal{I}_{[0,T']}<2(1-r)$
we put maximal duration of the \emph{off} intervals inside the unused
region $[T,t']$, making it identically zero, and the rest distribute
identically on intervals $[0,T]$ and $[t',T']$, so that all \emph{off}
intervals in $[0,T]$ will be multiplied by all \emph{off} intervals
in $[t',T']$. Then we have $C_{TA}(t',T')\leq\left(T-\left(T_{[0,T']}^{-}-(t'-T)\right)/2\right)/T=\mathcal{I}_{[0,T']}/(2(1-r))$,
where again, this upper bound is achievable. Considering the lower
bound, for $T_{[0,T']}^{-}>T$ or equivalently $\mathcal{I}_{[0,T']}<r$,
$C_{TA}(t',T')\geq0$ and can reach zero, because we can make the
whole interval $[0,T]$ zero. For $\mathcal{I}_{[0,T']}\geq r$ we
have $C_{TA}(t',T')\geq\left(T-T_{[0,T']}^{-}\right)/T=\left(\mathcal{I}_{[0,T']}-r\right)/(1-r)$.
Summarizing for $r\geq1/2$ we have Eq. (\ref{eq:Cta-bounds-r>1/2}).

The case of $r<1/2$, when $t'<T$, is more complicated. We note that
if the interval lies inside of $[t',T]$ it will be used twice, by
both functions $I(t)$ and $I(t+t')$. Therefore, to minimize $C_{TA}(t',T')$
for a given $T_{[0,T']}^{+}$ it seems desirable to put as much as
possible of the \emph{off} intervals into $[t',T]$. This is a good
idea until we can make these intervals to be multiplied by the \emph{on}
intervals. If there is too much \emph{off} time inside $[t',T]$ then
some zeros inside $[t',T]$ will necessarily multiply other zeros
inside $[t',T]$, the situation we want to avoid. This can happen
only if $r<1/3$. Therefore let us consider only the case of $1/3\leq r<1/2$.
Then if $T_{[0,T']}^{-}<T-t'$ or equivalently $\mathcal{I}_{[0,T']}>2r$
yields $C_{TA}(t',T')\geq\left(T-2T_{[0,T']}^{-}\right)/T=\left(2\mathcal{I}_{[0,T']}-r-1\right)/(1-r)$.
For $\mathcal{I}_{[0,T']}\leq2r$ we have $C_{TA}(t',T')\geq\left(T-2(T-t')-(T_{[0,T']}^{-}-(t-t'))\right)/T=\left(\mathcal{I}_{[0,T']}+r-1\right)/(1-r)$,
assuming that if this bound is negative it is replaced by 0. For the
upper bounds it follows that if $T_{[0,T']}^{-}\leq2[T-2(T-t')]$
or $\mathcal{I}_{[0,T']}\geq3-6r$ then $C_{TA}(t',T')\leq\left(T-T_{[0,T']}^{-}/2\right)/T=\left(\mathcal{I}_{[0,T']}+1-2r\right)/(2(1-r))$
and if $T_{[0,T']}^{+}\leq3(T-t')$ or $\mathcal{I}_{[0,T']}\leq3-6r$
then $C_{TA}(t',T')\leq\left(2T_{[0,T']}^{+}/3\right)/T=2\mathcal{I}_{[0,T']}/(3(1-r))$.
Summarizing for $1/3\leq r<1/2$ yields Eq. (\ref{eq:Cta-bounds-1/3<r<1/2}).

Finally, for small \emph{r} consider a simple counter-example showing
that the bound $C_{TA}(t',T')\leq\left(\mathcal{I}_{[0,T']}-r\right)/(1-r)$
can be overcome, in principle, for any $\mathcal{I}_{[0,T']}$. Let
$T'/t'=1/r$ be an integer. For any $T_{[0,T']}^{-}$ we then can
distribute the \emph{on} and \emph{off} times by first filling the
interval $[0,t']$ with \emph{on} time from 0 to $rT_{[0,T']}^{+}$
and filling the remainder (from $rT_{[0,T']}^{+}$ to $t'=rT'$) with
\emph{off} time. Rest of the intervals, $[t',2t']$, $[2t',3t']$,
..., $[T,T']$ are filled in exactly the same way. Then clearly $C_{TA}(t',T')=\left(rT_{[0,T']}^{+}(1-r)/r\right)/T=\mathcal{I}_{[0,T']}>\left(\mathcal{I}_{[0,T']}-r\right)/(1-r)$
for $\mathcal{I}_{[0,T']}<1$. The value $\mathcal{I}_{[0,T']}$ is
not an upper bound either, in general, as can be seen, e.g., from
Eq. (\ref{eq:Cta-bounds-1/3<r<1/2}).


\begin{thebibliography}{10}
\bibitem{Allegrini-wrong}P. Allegrini, P. Grigolini, L. Palatella and B. J. West, \emph{Phys.
Rev. E} \textbf{70} 046118 (2004).
\bibitem{NadlerStein91}W. Nadler and D. L. Stein, \emph{Proc. Natl. Acad. Sci. USA} \textbf{88},
6750 (1991). 
\bibitem{GoychukHanggi02}I. Goychuk and P. H\"{a}nggi, \emph{Proc. Natl. Acad. Sci. USA} \textbf{99}
3552 (2002). 
\bibitem{Dewey02}T. G. Dewey, \emph{Drug Discovery Today} \textbf{7} S170 (2002).
\bibitem{Roy}S. Roy, I. Bose and S. S. Manna, \emph{International J. Modern Phys.
C} \textbf{12}, 413 (2001).
\bibitem{Masuda}N. Masuda and K. Aihara, \emph{Neural Computation} \textbf{15} 1341
(2003). 
\bibitem{Korobkova04}E. Korobkova, T. Emonet, J. M. G. Vilar, T. S. Shimizu and P. Cluzel,
\emph{Nature} \textbf{428}, 574 (2004).
\bibitem{Haase04}M. Haase, C. G. H\"ubner, E. Reuther, A. Herrmann, K. M\"ullen and
Th. Basch\'e, \emph{J. Phys. Chem. B} \textbf{108}, 10445 (2004).
\bibitem{Nirmal}M. Nirmal, B. O. Dabbousi, M. G. Bawendi, J. J. Macklin, J. K. Trautman,
T. D. Harris, L. E. Brus, \emph{Nature} \textbf{383} 802 (1996). 
\bibitem{Kuno}M. Kuno, D. P. Fromm, S. T. Johnson, A. Gallagher and D. J. Nesbitt,
\emph{Phys. Rev. B} \textbf{67} 125304 (2003).
\bibitem{Ken}K. T. Shimizu, R. G. Neuhauser, C. A. Leatherdale, S. A. Empedocles,
W. K. Woo and M. G. Bawendi, \textit{Phys. Rev. B} \textbf{63} 205316
(2001). 
\bibitem{Dahan}G. Messin, J. P. Hermier, E. Giacobino, P. Desbiolles and M. Dahan,
\emph{Optics Letters} \textbf{26} 1891 (2001).
\bibitem{Brokmann}X. Brokmann, J. P. Hermier, G. Messin, P. Desbiolles, J.-P. Bouchaud,
and M. Dahan, \emph{Phys. Rev. Lett.} \textbf{90} 120601 (2003).
\bibitem{Zumofen04}G. Zumofen, J. Hohlbein and C. G. H\"ubner, \emph{Phys. Rev. Lett.}
\textbf{93} 260601 (2004).
\bibitem{GL}C. Godr\`{e}che and J. M. Luck, \emph{J. Stat. Phys.} \textbf{104}
489 (2001). 
\bibitem{Bald}A. Baldassarri, J. P. Bouchaud, I. Dornic, and C. Godr\`{e}che \emph{Phys.
Rev. E} \textbf{59} R20 (1999).
\bibitem{BouchaudGeorges90}J-P. Bouchaud and A. Georges, \emph{Physics Reports} \textbf{195}
127 (1990).
\bibitem{Schlesinger}J. Klafter, M. F. Shlesinger, and G. Zumofen, \emph{Phys. Today} \textbf{49}
33 (1996).
\bibitem{MetzKlaf00}R. Metzler and J. Klafter, \emph{Physics Reports} \textbf{339} 1 (2000).
\bibitem{BJS04}E. Barkai, Y. Jung and Silbey, \emph{Annu. Rev. Phys. Chem.} \textbf{55}
457 (2004).
\bibitem{ZK}G. Zumofen, and J. Klafter, \emph{Phys. Rev. E} \textbf{47} 851 (1993). 
\bibitem{Marinari93}E. Marinari and G. Parisi, \emph{J. Phys. A} \textbf{26} L1149 (1993).
\bibitem{Bouchaud92}J. P. Bouchaud, \emph{J. Phys. I France} \textbf{2} 1705 (1992). 
\bibitem{Cheng}E. Barkai, and Y. C. Cheng \emph{J. Chem. Phys.} \textbf{118} 6167
(2003). 
\bibitem{Chaos}E. Barkai \emph{Phys. Rev. Lett.} \textbf{90} 104101 (2003).
\bibitem{MB_JCP}G. Margolin and E. Barkai, \emph{J. Chem. Phys.} \textbf{121} 1566
(2004).
\bibitem{LineShape04}G. Aquino, L. Palatella and P. Grigolini, \emph{Phys. Rev. Lett.}
\textbf{93} 050601 (2004).
\bibitem{Verberk}R. Verberk, and M. Orrit, \emph{J. Chem. Phys.} \textbf{119} 2214
(2003).
\bibitem{MB_PRL}G. Margolin and E. Barkai, \emph{Phys. Rev. Lett.} \textbf{94} 080601
(2005).
\bibitem{1<theta<2}For a power law $\psi(\tau)$ with $1<\theta<2$, the distribution
of correlation function will also approach the Dirac delta, because
the mean sojourn time is finite. However, the variance of this distribution
will go to zero as $(T')^{1-\theta}$, as opposed to $(T')^{-1}$
for $\psi(\tau)$ with finite second moment.
\bibitem{Lamp}J. Lamperti, \emph{Trans. Amer. Math. Soc.}  \textbf{88} 380 (1958). 
\bibitem{Feller}W. Feller, \emph{An Introduction to Probability Theory and its Application}
Vol. 2, Wiley New York (1970).
\bibitem{Dhar}A. Dhar and S. N. Majumdar, \emph{Phys. Rev. E} \textbf{59} 6413 (1999).
\bibitem{BelBarkai}G. Bel and E. Barkai, cond-mat/0502154 (2005).
\bibitem{Majumdar}S. N. Majumdar, cond-mat/9907407 (2004).
\bibitem{Pelton04}M. Pelton, D. G. Grier and P. Guyot-Sionnest, \emph{Appl. Phys. Lett.}
\textbf{85} 819 (2004).
\bibitem{Zumofen93}G. Zumofen and J. Klafter, \emph{Physica D} \textbf{69} 436 (1993).
\bibitem{Davidsen}J. Davidsen and H. G. Schuster, \emph{Phys. Rev. E} \textbf{65} 026120
(2002).
\bibitem{Schriefl05}J. Schriefl, M. Clusel, D. Carpentier and P. Degiovanni \emph{Europhys.
Lett.} \textbf{69} 156 (2005), and cond-mat/0501301 (2005).
\bibitem{Xie}H. Yang, G. Luo, P. Karnchanaphanurach, T.-M. Louie, I. Rech, S. Cova,
L. Xun and X. S. Xie, \emph{Science} \textbf{302} 262 (2003).
\bibitem{Weitz}I. Y. Wong, M. L. Gardel, D. R. Reichman, E. R. Weeks, M. T. Valentine,
A. R. Bausch and D. A. Weitz \emph{Phys. Rev. Lett.} \textbf{92} 178101
(2004).\end{thebibliography}
\end{document}